# The New Anticipatory Governance Culture for Innovation: Regulatory Foresight, Regulatory Experimentation and Regulatory Learning[*]

**Abstract**
With the rapid pace of technological innovation, traditional methods of policy formation and legislating are becoming conspicuously anachronistic. The need for regulatory choices to be made to counter the deadening effect of regulatory lag is more important to developing markets and fostering growth than achieving one-off regulatory perfection. This article advances scholarship on innovation policy and the regulation of technological innovation in the European Union. It does so by considering what building an agile yet robust anticipatory governance regulatory culture involves. It systematically excavates a variety of tools and elements that are being put into use in inventive ways and argues that these need to be more cohesively and systemically integrated into jurisdictions' regulatory toolbox. Approaches covered include strategic foresight, the critical embrace of iterative policy development and regulatory learning in the face of uncertainty and the embrace of bottom-up approaches to co-creation of policy such as Policy Labs and the testing and regulatory learning through pilot regulation and experimentation. The growing use of regulatory sandboxes as an EU policy tool to boost innovation and navigate regulatory complexity as seen in the EU AI Act is also probed.

## 1   Introduction

The speed of advances in the innovation age throws up both regulatory ambiguity for existing regulation and multifaceted challenges for new policy design. Regulators need to ensure that they are not outpaced by innovation thereby impeding responsible market development. But should they not be aiming higher? How can they anticipate what is coming next? This means asking how do we flip the problem of regulatory lag in the sphere of technological innovation so that rather than trying to merely keep up, continuous processes of regulatory foresight and experimentation become integral to the solution, transforming the business of regulatory policy formation and fundamentally altering its dynamic. This article contends that policymakers need to go beyond keeping pace with what has emerged. Instead they need to embrace an imaginative, forward-driven, anticipatory governance culture that proactively anticipates innovation and engages with it head on. Anticipatory governance can lead to bold new and imaginative approaches to policy design; combining both evidence and informed speculation about what the future holds and how policy should shape it. Anticipatory governance for innovation provides the tools to address the lag caused by innovation outpacing regulation and policy. Furthermore, this type of future-thinking regulatory mindset can help to proactively foster innovation and catalyse the development of new markets for it. That said, assuming that new policy actions or regulatory interventions are an inevitable immediate outcome of foresight processes is to misconceive the full continuum of approaches within which anticipatory governance operates and the relevance of an assessment of readiness. Anticipatory governance is timely but not hasty. Reflecting that, in the first instance anticipatory governance foresight functionality may usefully yield a direction of resources and capacity-building around early emerging innovation and potential policy issues.

---

[*] Professor Deirdre Ahern , Professor in Law, Technologies, Law & Society Research Group, Trinity College Dublin.

Reflecting the status quo on the ground where anticipatory governance is new and far from firmly established, scholarship on anticipatory governance including policy labs and other forms of regulatory learning is still very much at an emerging stage. In discussing how innovation policy and regulation is made, this article advances the literature and contributes to debate on innovation policy. Anticipatory governance can be considered qua meta-regulation, meta-regulation being used here in Peter Grabosky's sense of the study of regulation, not just as something that applies to certain actors, but also as a system (applied to a sub-field of regulation or policymaking) that can itself be studied in its own right.[1]. Here the aim is to give a systems view of anticipatory governance while also drilling down into its constituent parts.

An anticipatory governance approach calls for a mindset of considerable openness to taking calculated risks in trying timely fresh approaches to devising regulatory policy for innovation, rather than simply relying on tried and tested but less than agile routes to policy development. Breaking down anticipatory governance, it is postulated here that three key elements form the triumvirate of anticipatory governance – regulatory foresight, regulatory experimentation and regulatory learning. Calling for the embedding of a robust anticipatory governance culture around innovation, this article unpacks how new experimental policy modalities are beginning to manifest and considers their suitability to tackle future-planning for innovation. This requires a regulatory vision that embraces future-scanning, collaborative iteration and participatory policy design where the process of policymaking is not viewed as linear. Furthermore, leaving behind top-down regulation, new actors can come to the fore in bottom-up construction of the way forward and its policy dimensions as experimental and creative methods are used to imaginatively and iteratively envision versions of the future and their consequences.

It is understandable that policymakers and regulators coming from a traditional culture of top-down regulation worry about getting it wrong in taking a stand on technology, but this is not the mindset of anticipatory governance which focuses on direction and collaboration not perfection. The direction of travel should be towards an anticipatory governance culture that is not simply focused on short-term predictability but is comfortable with probing what Pólova and Nascimento term 'tangible speculative futures'.[2] As the Organisation for Economic Co-operation ('OECD') notes, '[s]trategic foresight is a structured and systematic way of using ideas about the future to anticipate and better prepare for change. It is about exploring different plausible futures that could arise, and the opportunities and challenges they could present. We then use those ideas to make better decisions and act now.'[3] Experimentation designed to facilitate innovation such as regulatory sandboxes and trialling pilot regulations can yield insights that provide huge regulatory learning value to regulators around the efficacies and gaps in regulation. The advantages of such small-scale experimentation are the rapid feedback gained while also limiting the cost of both regulatory and innovator failure.[4]

Section 2 begins by introducing the widening regulatory gap that innovation brings, leading to the regulatory, market and societal problem of regulatory lag. From there Section 3 gives an exposition of the elements needed to build a robust anticipatory governance regulatory culture that keeps pace with innovation. This section draws on developments that are taking place in increasing pockets around the globe in order to identify the key elements of an anticipatory governance regulatory culture. Section 4 introduces the concept of regulatory foresight–the first element of the anticipatory governance triumvirate–while Section 5 discusses iterative policy development as a means of refinement and iteration amid uncertainty. From there we explore in Section 6 the emergence of regulatory experimentation, the second element of the anticipatory governance triumvirate. Relevant experimentation mechanisms explored include regulatory sandboxes, pilot regulation, policy labs and experimentation clauses. These constitute collaborative, creative mechanisms for policy development, not only to boost innovation, but to generate regulatory learning, the third element of the anticipatory governance triumvirate. Regulatory learning as a practice is further explored in Section 7. Section 8 confronts some challenges for anticipatory governance as it emerges including the question of how to evaluate the effectiveness of anticipatory governance in a given system. The Conclusion provides some final thoughts concerning the contribution and future evolution of anticipatory governance as a culture to proactively anticipate and engage with the challenges presented by innovation.

---

[1] Grabosky (2017). See also Morgan (2003). Meta regulation, has, for example, been studied in the context of a better regulation agenda around public health: Lauber and Brooks (2023).
[2] Pólova and Nascimento (2021), p. 2.
[3] Organisation for Economic Co-operation and Development (2023).
[4] Lee and Ling (2020).

## 2      The Widening Regulatory Gap That Innovation Brings

Well-timed, proportionate regulation for innovation represents the Holy Grail. In an unregulated market moral hazard applies and detrimental societal impacts may result from regulatory gaps which highlight the costs of under-regulation. Meanwhile regulatory uncertainty around how a pre-existing regulatory framework applies to innovation adds to the costs of market entry. Over-regulation may deter market entry due to the regulatory barriers to entry being too high as compared with other markets. Furthermore, some forms of regulation attract various negative labels such as red tape, regulatory burden, over-regulation or 'juridification'.[5] Calibrating regulation for innovation is thus inherently difficult.

Gaps in regulation create uncertainty which operates as an invisible but significant roadblock hampering the normal functioning of the innovation cycle, often negatively impacting through immeasurable foregone opportunities of societal benefit from commercialisation and associated wealth creation that would otherwise have occurred. However, the impacts of regulatory gaps may not fall uniformly: when regulatory lag occurs, legacy regulatory frameworks may favour incumbents.[6] Arbitrage is the trading phenomenon of buying and selling currencies, equities or commodities, leveraging favourable price differences across markets. Similarly, regulatory lag across jurisdictions may be exploited for gain through the phenomenon of regulatory arbitrage.[7] Regulatory arbitrage occurs where firms seek to capitalise on differing levels of regulation across states in order to site their business in a location which offers the least onerous regulatory framework, thereby reducing regulatory costs. Market fragmentation is a typical problem in the European Union, enabling location selection choices by market entrants among regimes with differing levels of market intervention. The game of regulatory arbitrage can be substantially reduced where maximum regulatory harmonisation frameworks exist for innovation, as highlighted by the EU AI Act.[8] However, in many areas no such framework exists and policy formulation processes are drawn out.

We are faced with a widening 21st century regulation gap around technological innovation as regulation is unmatched by the sheer pace and transformative impact of disruptive innovation. This matters for the robustness and agility of a legal system but it also has considerable social and economic ripples because innovation drives social and economic progress. An inescapable truth is that innovation, be it nanotechnologies, blockchain, Internet of Things or the transformative power of Artificial Intelligence ('AI'), is radically disrupting the worlds of healthcare, manufacturing and finance to name but a few. Robotics are automating manufacturing and assembly processes. Technological change is constantly driving and refining new business models as seen in AI-driven fintech and markets in DLT-based cryptocurrencies which shake up traditional banking and investment markets, bringing both potential benefits and risks.[9] New fintech business models have arrived such as peer-to-peer lending, banking as a service and Buy Now, Pay Later (a trending payment model where consumers are offered the opportunity to order and receive goods up-front and pay for them in instalments, conveniently avoiding existing lending rulebooks).[10]

---
[5] Teubner (1998), pp 389-440.
[6] Ahern (2018); Clements (2023).
[7] Pollman (2019); Ahern (2021).
[8] European Commission (2021).

The speed of innovation adds immeasurably to the regulatory challenge. This is where policymakers and regulators flounder as the pacing problem leads to regulation that is either too hot or too cold in temperature rather than about right. The quandary is that regulating too early may stifle the innovation lifecycle and deter investment needed to grow innovation. By contrast, delaying and regulating too late risks investments being stranded.[11] The pacing problem also brings to the fore the so-called 'Collingridge Dilemma'[12] whereby to wait too long to regulate is to let technology take hold in society and the economy, making exerting subsequent control over it very difficult. The evolution of BigTech platform models is a case in point. Then there is the challenge of devising proportionate regulation that is neither too lenient nor too harsh. Timing and regulatory content are thus complex variables for policymakers to confront.

Some states may be content to play a waiting game to observe what leader states (or the EU) does rather than risk their efforts being viewed as ill-judged. A passive 'wait and see' approach is one form of under-regulation. One step up is 'light touch' regulation through low intervention. This may encourage initial market entry by innovators due to low regulatory costs for market entry. However, technological innovation is never fully fathomable and a hands-off regulatory approach that fails to continually fully engage with risks inherent in emerging business models may come acropper. The collapse of the FTX trading platform for cryptocurrencies in late 2022 and founder Sam Bankman-Fried's subsequent conviction for fraud and money laundering[13] made plain to many regulators that leaving the crypto sector to operate outside the regulatory perimeter was no longer an option, not least given the potential impacts on financial stability.[14]

The stakes in regulating innovation are very high. Yet at the same time there are many unknowns for policymakers and regulators exploring unfamiliar regulatory territory in relation to new technologies, their uses and unpredictable effects on society and the economy. All of this helps to explain the widening regulatory gap that innovation brings. In the words of the acclaimed regulatory scholar, John Braithwaite, ['r]egulation is hard to do well – politically hard, technically hard, legally hard'.[15] Devising regulation brings with it a dynamic tension: aiming to reduce risk comes with the possibility that public benefit may also be cut down in both foreseen and unforeseen ways. There is also an additional danger that complexities and uncertainties around innovation may cause policymakers to 'pivot according to political winds.'[16] This has been seen with regulatory volte faces around crypto-assets.[17] In 2023 the European Union was the first major jurisdiction to regulate crypto-assets in the form of the Markets in Crypto-Assets Regulation ('MICA'),[18] a key deliverable of the Digital Finance Strategy.[19] Yet, prior to 2017, jurisdictions globally saw Distributed Ledger Technologies as too immature in their potential applications to financial markets consider regulating.[20] Regulators were somewhat on the backfoot as it was only when the markets erupted with Initial Coin Offerings that a tipping point catalysed and they began to realise that they needed to consider regulatory opportunities and risks.[21]

---

[9] Distributed Ledger Technologies ('DLT').
[10] Gerrans, Baur and Lavagna-Slater (2022); Ahern (2023).
[11] Regulatory Horizons Council (2022), p. 7.
[12] Collingridge (1980), p. 19.
[13] Miller J and Oliver J, 'Sam Bankman-Fried convicted of fraud over FTX collapse' *Financial Times* 3 November 2023 https://www.ft.com/content/24d153b0-0c28-4946-acbe-2e93329bca52. Accessed 12 March 2024.
[14] That has not simplified the issue of how to classify and regulate it.
[15] Braithwaite(2008), p. 62.
[16] Tõnurist and Hanson (2020), p. 13.
[17] Arner et al. (2023).

The above discussion illustrates the multi-faceted complexities associated with regulating innovation. For present purposes the focus is on one specific angle which boils down to the regulatory difficulty of anticipating what comes next with innovation and its applications. Failure to anticipate what is coming down the tracks exponentially expands the widening regulatory gap that already exists around new and disruptive innovation. There is a certain inevitability about anticipation failure around innovation where policymaking methods deployed are centralised, traditional and inflexible and thus non-agile and suboptimal at addressing existing innovation that is available, never mind emergent innovation that is in flux, nascent innovation and stepping up to the mega challenge of anticipating what comes next. In determining how to approach innovation, the content of regulation is non-trivial. But the meta-regulation question which this article is invested in is 'the how' rather than the 'what' - how in a given regulatory system regulatory responses to innovation can be temporally arrived at in a manner that overcomes the pacing problem. This fundamental question therefore presents a problem of process and method, not just substance. As Nesta, an influential UK innovation agency has declared, the challenges brought by innovation 'require new regulatory practices and stances, not just (or always) new regulatory initiatives or bodies.'[22] Given the limitations associated with existing policy and regulatory processes to really act as a positive catalyst of events by taking the bull by the horns around innovation to stay ahead of the curve, states ought to be investigating embedding the building blocks of an anticipatory regulatory governance culture.

## 3    Building a Robust Anticipatory Innovation Governance Regulatory Culture

In jurisdictions where anticipatory governance is not in place, policymakers are typically briefed on developments that are already market-ready or on the market. That scenario is a back foot, reactive one for policymakers rather than a proactive one. To be responsive rather than reactive to technological innovation we need a form of regulatory culture that leans into the inevitable uncertainty around how technology will evolve, its use and the risks it presents, rather than choosing to simply stand back and wait for it to become more established. Anticipatory governance is defined by the OECD as the 'systematic embedding and application of strategic foresight throughout the entire governance architecture, including policy analysis, engagement, and decision-making.'[23] Foresight is central. In line with that, jurisdictions with such a culture will be more likely to have understood the potential trajectory of innovation at a much earlier stage and potentially to have dipped their toe into policymaking waters in one form or another. This is because a systematic anticipatory innovation governance culture looks down the line at near future scenarios and probes them thoroughly. In essence, an anticipatory regulatory governance culture allows a country to first, proactively look down the horizon and second, to respond in a nimble fashion including through identifying new priorities and harnessing inventive regulatory craft to iterate anticipatory governance. Intrinsic to this is a regulatory culture of experimentation and iteration. Thus far the anticipatory governance movement remains at an early stage--anticipatory governance has been embedded in a few countries for years but is now on the cusp of breaking into the mainstream. Countries that have been pioneers and are leading out in this sphere include Australia, Canada, Estonia, Finland and Singapore. Jurisdictions averring to be advocates of agile governance are often in the early stages. They may be 'talking the talk' but 'walking the walk' and the achievement of tangible outputs takes time. Assisting with this, best practices are beginning to being disseminated in networks such as the OECD's Government Foresight Community[24] and the European Commission's EU-wide Foresight Network.[25] Meanwhile pockets of scholarship[26] across disciplines and policy reports[27] on anticipatory governance have been slowly building over the last number of years. With the growth in interest in how to confront the pacing problem for new technologies, it is timely to advance the conceptualisation and understanding of the new anticipatory governance as an adaptive solution to both keeping pace and winning at regulation.

---

[18] Markets in Crypto-Assets Regulation (EU) 2023/1114 ('MICA').
[19] European Commission (2020).
[20] Ferreira and Sandner (2021).
[21] Ferreira and Sandner (2021).
[22] Armstrong et al. (2019), p. 5.
[23] Organisation for Economic Co-operation and Development (2019), p. 3.

Taxonomically, it is posited here that three elements (discussed in detail in subsequent sections) come together to form the triumvirate of anticipatory governance – regulatory foresight, regulatory experimentation and regulatory learning. An anticipatory governance approach involves embedding methods into regulatory culture for thinking not only about regulating new or emerging technologies but also about the opportunities and potential applications (as well as the risks) of emergent technologies that have not fully materialised yet. This starts with the right mindset. Anticipatory governance at its best is supply-led, proceeding based on intelligence gathered from foresight activities. As regards governance, reflecting its anticipatory nature, anticipatory governance does not simply wait for policy gaps to open up, it boldly leads. Policymakers should 'not only forecast, but also prepare for and respond'.[28] The Trojan princess Cassandra suffered the fate of being able to see the future but being ignored. If this occurs in relation to what foresight activities portend, it is a waste of resources.

Anticipatory governance legitimises a capacity-building and policy focus beyond the present or immediate future. Resources are devoted to a deep dive imagining of what a future or alternative futures could look like with contributions made by innovation advances. For instance, within the deep tech space, quantum computing, although far from reaching its intended potential, is getting closer every day and the science is beginning to mature such that its technological applications in the next few years can now be envisaged. In framing anticipatory governance below as a new style of regulatory culture, it is important to emphasise that regulatory agility involves a smart flexibility of approach. Quick fire policy changes in complex fields of innovation that are still in the early stage of evolution are by no means frequent, nor should they be. Policymakers still need to look before they leap and to invoke the precautionary principle as needed. But as detailed below, where appropriate, a foresight-informed, iterative approach to policymaking can be activated to be bolder than a traditional incremental one whereby policy is marginally adjusted gradually over time.

As such, the aims of regulation within anticipatory governance can be bigger because adjustments and fine-tuning can be more rapidly undertaken in response to operational signals received. Within this policy agenda, policymakers and the public sector participate in new methods. Iterative policymaking shows the public sector responding to the impetus of technological innovation with innovation of their own, an occurrence that falls neatly within the Schumpeter thesis of innovation as a process of creative destruction.[29] The way in which innovation policy and regulation is iterated within anticipatory governance chimes with the contemporary understanding in regulatory theory that regulation is layered with 'multiple levels and sources'.[30] Thus an anticipatory governance culture is comfortable with elements of polycentric governance and networked governance being at play.[31]

---

[24] https://www.oecd.org/strategic-foresight/ourwork/. Accessed 12 March 2024.
[25] https://commission.europa.eu/strategy-and-policy/strategic-planning/strategic-foresight_en#eu-wide-foresight-network. Accessed 12 March 2024.
[26] Fuerth and Faber (2013); Cagnin et al. (2013); Guston (2014); Kimbell and Bailey (2017); Kwek and Parkash (2020).
[27] Organisation for Economic Co-operation and Development (2019); Kert et al. (2022); European Commission (2023d).
[28] ElZarrad et al. (2022), p. 1333.

The rationale for building a robust, anticipatory governance regulatory culture has both public interest and economic roots. Parker and Braithwaite famously described regulation in terms of its power to 'influenc[e] the flow of events'.[32] This resonates here. Outcomes of anticipatory governance do not only relate to pre-emptive risk mitigation. Regulatory foresight activities can drive well-calibrated regulatory policy initiatives on the supply-side that promote innovation and attract innovators and investors. The resulting market-building benefits qualify supremely as influencing the flow of market events. Such considerations are evident in the UK Financial Conduct Authority's decision to propose regulating fiat-backed Stablecoins with a view to harnessing a market for early-adopters while mitigating potential risks.[33] Regulation within an anticipatory governance culture is thus not only concerned with fixing market failures, but also about anticipating and facilitating the creation of new markets for innovation and delivery of societal benefits.[34] An anticipatory governance approach is also consistent with the perspective that regulation is not just about risk reduction, it is also about 'governments' need to demonstrate their mastery over uncertainty'.[35]

While the justification for anticipatory governance for innovation is clear, how should an anticipatory governance culture be rolled out? Should an anticipatory governance culture be formally framed within a new set of provisions? This has not typically been the case in countries that have begun to embrace it. Arguably the various elements of what constitutes an anticipatory governance culture do not need to be framed into hard law. Doing so could institute a counter-productive, rigid straitjacket that would be too restrictive, particularly given the early stage of evolution of anticipatory governance. Moreover, the tenets of anticipatory governance are more about attitudes than rules. By contrast, a set of flexible guiding principles or a soft code of practice with general principles around anticipatory governance could be helpful in inculcating this culture. To that end, five key guiding tenets of an agile, anticipatory governance culture are posited here as follows:

(i) Anticipatory governance work begins with an open mind as to both the opportunities and the risks of a potential innovation or application of one and how they balance each other;
(ii) Anticipatory governance foresight activities embrace new forms of inputs and evidence to evaluate potential futures and their impacts;
(iii) In response to an anticipated innovation stimulus, anticipatory governance evaluates what if any policy response is merited and when;
(iv) Anticipatory governance looks to create policy that works for cutting-edge innovation through trialling iterative approaches to regulatory policy development that allow for mutual regulator-stakeholder learning;
(v) Anticipatory governance is a reflective practice that is open to incorporating insights from regulatory learning and new information on innovation capabilities and application.

---

[29] Schumpeter (2008).
[30] Drahos and Krygier (2017), pp 6-7.
[31] McGann et al. (2018).
[32] Parker and Braithwaite (2003), p. 119.
[33] Financial Conduct Authority (2023a).
[34] Prosser (2006).
[35] Haine (2017), p. 192.

Even if the rationale for anticipatory governance and its key tenets are clear, laying the right groundwork requires meticulous attention to detail in terms of structures, interactions and decision-making. Building a strong anticipatory governance regulatory culture necessitates intentional and focused work to develop the right structures, approaches and resources. A good regulatory foresight unit firing on all cylinders is by nature resource-heavy. Building an anticipatory governance regulatory culture which enshrines not just horizon-scanning but collaborative, inter-disciplinary iterative approaches to policy development requires thinking about public service architecture. Furthermore, while an anticipatory governance culture has the potential to deliver imaginative strategies, policies and regulation, this type of culture is new and needs to be robustly embedded in a governance system with end to end buy-in for beneficial impacts to occur.

An obvious challenge which presents here is organisational cultural and technological readiness to embrace anticipatory governance for innovation.[36] The culture of governmental policymaking may prove resistant to the requisite shake-up needed to integrate a holistic anticipatory regulatory governance culture. Achieving a networked, whole-of-government approach undoubtedly requires bold commitment that is values-led and committed rather than tokenistic. Certainly this culture requires new ways of working and new expertise and skills within policy teams. People working on policy in government departments and regulators will not necessarily have relevant technical or industry expertise to work with the cutting edge of technological innovation. Therefore targeted recruitment and capacity-building to learn and institutionalise foresight and other anticipatory governance systems is vital. In assessing how dynamically an anticipatory governance system performs and how fluidly its constituent parts interact, policymakers and regulators who have received appropriate training from foresight practitioners are likely to be better enabled to work in new ways than those who have not and thus may resist full buy-in. Strong networks and high level champions can a make a difference.[37] For example, in the United Kingdom, the Digital Regulation Cooperation Forum provides a forum for information sharing on digital regulatory matters and collaborative research and sharing of insights to ensure a cohesive approach to regulatory policy. Change in how the public sector operates is certainly capable of occurring.[38] Already technological innovation is influencing how policymaking, regulation and regulators function using code and data-driven models and regulatory technology ('RegTech').[39] RegTech is not just driving enhanced firm-level compliance, the term is also being used to describe how big data analytics tools can be deployed to powerfully aid regulators and supervisors ('SupTech') in their monitoring and enforcement roles. Furthermore, a dramatic digital transformation of the public sector across the European Union is underway.[40]

Getting the architecture right for anticipatory governance is pivotal to building the right foundation for the whole endeavour. This architecture should be procedurally formalised so that anticipatory governance is optimally institutionalised and resourced. To build a robust anticipatory governance culture for innovation, ideally a national policy office should lead out on anticipatory governance. It needs to be close to the scientific frontier of research and, where appropriate, act as a source of funding to develop the innovation ecosystem. As identified by Fuerth, anticipatory governance requires (1) a foresight system; (2) a networked system for integrating foresight into the policy process; (3) a feedback system on performance; and (4) an institutional culture that is open-minded.[41] In line with this, anticipatory governance is likely to work best where there is a coordinated effort at government level with inclusive and open cross-government participation to build in inter-disciplinarity as seen in the UK's civil service Policy Lab[42] which works across government. Foresight work may be optimised where it is carried out at some remove, that is with some independence from political power.[43] Indeed, a joined up approach mediated by a central agency in a jurisdiction with some independence from central government may lead to both a more boundary-pushing and cohesive approach that avoids individual ministries having to unnecessarily reinvent the wheel. In terms of costs and joined up thinking, duplication of foresight efforts in separate specialised departments should be avoided.

---

[36] Bolton and Mintrom (2023).
[37] Organisation for Economic Co-operation and Development (2021), p. 9.
[38] On the planned use of AI in government in the United Kingdom see Fisher L, 'UK Government to Trial "Red Box" AI tools to Improve Ministerial Efficiency' *Financial Times* 28 February 2024 https://www.ft.com/content/f2ae55bf-b9fa-49b5-ac0e-8b7411729539. Accessed 12 March 2024.
[39] Micheler and Whaley (2019); McNulty et al. (2022).
[40] European Commission (2022a).The culmination of this vision is the Interoperable Europe Act on which political agreement was reached on 14 November 2023.

Within the European Union, strategic foresight is led out under the direction of the Vice-President of the European Commission with implementation by the Secretariat-General and the Joint Research Centre. Strategic foresight is networked by means of the Strategic Foresight Network between all Directorates-General. There is also a networked effect through the EU-wide Foresight Network that draws on Member States' foresight capacities. Fostering innovation in policymaking is an avowed priority objective of the European Commission's Joint Research Centre.[44] It is embedding foresight activities across all its activities to strengthen a culture of anticipation around emerging issues.[45] An excellent example is provided by the EU Policy Lab which operates within the Foresight, Modelling, Behavioural Insights and Design for Policy Unit of the Joint Research Centre. This Unit is tasked with providing advice that is evidence-based and works from strategic foresight to policy design to support the European Commission in setting its policy programme.[46] A networked approach is provided by the EU Policy Lab within the Joint Research Centre which delivers evidence-based policy insights. The Joint Research Centre views it as 'both a physical space and a way of collaborative working that combines foresight, behavioural insights and design for policy to explore, connect and find solutions for better policies.'[47]

Seeing the big picture is easier with a coordinated inter-agency approach. Rather than activity occurring in isolated pockets, there is room for cross-pollination of ideas and expertise and shared capabilities. Perhaps the most networked example comes from world leader, Singapore, which has had a Risk Assessment and Horizon Scanning Experimentation Centre since 2007. The Centre for Strategic Futures in the Prime Minister's Office provides the central node for future work while a technology platform enables collaboration across agencies. In Canada, a country that is very strong in this field, Policy Horizons Canada acts as a centre of excellence in foresight that helps to support the Government with future-oriented decision-making in order to make better policy decisions. Canada's Centre for Regulatory Innovation champions a whole-of-government approach to regulatory experimentation and removal of barriers to innovation including in the cleantech sphere.

In the sections that follow, key operational aspects of anticipatory governance are unpacked. Before doing so, it is worth underlining that a successful anticipatory governance culture must build and maintain trust within the public sector, government and the wider public. This is an inclusive vision of anticipatory governance that is focused on the public interest as well as nurturing innovation.[48] Legitimacy within the wider population for anticipatory governance policy-related activities will be assisted by early and varied engagement around contested issues. Trust will also be enhanced by transparency and a willingness to put building blocks in place to manage legal and ethical risks while enabling innovation.[49] Participatory policymaking involves processes that go well beyond mere stakeholder consultation. Potential futures are explored and potential policy outcomes collaboratively arrived at, for example, through diverse participation in policy labs. There is space here for regulators, innovators and citizens to work together. Policy labs rely on tapping into the variety of expertise and ways of thinking in the room and creating an atmosphere of co-creation. This requires a delicate balance. Government should be careful not to cede the regulatory reins while being stakeholder-inclusive in constructing regulatory insights. It is well known in regulatory theory that where there is close cooperation between a regulator and a firm independence can take a back seat.[50] Co-creation is one thing, regulatory capture is another. Asymmetries of expertise and market awareness could lead to policymakers becoming dependent on industry stakeholders and not robustly considering whether their policy ideas are objectively beneficial. This could be countered by accommodating diverse, multi-stakeholder participation including giving a seat at the table to public interest representatives. A challenge for anticipatory governance concerns how to successfully manage the ability to have a strong evidence base with the need to also satisfy stakeholder inclusion in co-creating policy design.[51] Experimentation and pilot regimes may be perceived as having an exclusionary effect if invitations to participate are restricted. As Ranchórdas notes, this may be justified in scaled-down evidence based experiments but not where an experiment is poorly designed and unlikely to deliver evidence-based policymaking.[52] This could lead to distrust in other parts of government administration and among the wider community and business in relation to policy outcomes arrived at. This makes transparency and public accountability crucial in anticipatory governance.

---

[41] Fuerth (2009), p. 20.
[42] Countries where government is more devolved and less centralised such as Germany may not be able to realise a highly centralised model.
[43] Organisation for Economic Co-operation and Development (2021), p. 9.
[44] McAleavey (2023), p. 16.
[45] Heuer et al. (2022), p. 5.
[46] Pólova and Nascimento (2021).
[47] McAleavey (2023), p. 21.
[48] Macnaghten (2021).

## 4      Regulatory Foresight

Foresight obtained through forecasting is the apex component of an anticipatory governance regulatory culture for innovation. As defined by Fuerth, foresight is:

> 'the capacity to anticipate alternative futures, based on sensitivity to weak signals, and an ability to visualize their consequences, in the form of multiple possible outcomes. It is a means to visualize, rehearse and then refine in the mind, actions that would otherwise have to be tested against reality, where the consequences of error are irrevocable.'[53]

Within foresight, the practice known as 'horizon-scanning' is a systematic approach to move beyond the status quo to detecting early signs of new changes including disruptors that may develop into long-range market trends and making sense of them in order to anticipate their plausibility and potential future impact in order to generate strategies to prepare accordingly. The outputs of scanning events and trends are mapped. It is not just a job for policymakers but also for regulators who must keep an eye on market developments and Research and Development ('R&D') to consider what the implications would be if an important development such as a new form of financial innovation popped up in the future that fell outside the regulatory perimeter. Engagement with experts through interviews, surveys and workshops can help to weight and filter signals for significance. Indeed, future studies is a field that has engaged with systematic methodologies of predicting futures. Futures and foresight techniques can include driver mapping and scenario analysis to monitor changes in the technology landscape.[54]

---

[49] Kert et al. (2022).
[50] Drahos and Krygier (2017), p. 5.
[51] Bunea and Chrisp (2022).
[52] Ranchordás (2021), p. 14.
[53] Fuerth (2009), p. 16.

Regulatory foresight depends on gathering high quality intelligence and maintaining excellent points of intersection for dialogue between science, industry and policymakers in order to build a prescient vision for response in the form of strategies and policy responses. Once the signals picked up have been branded as significant, scenario planning is an interactive process that is followed by envisioning a roadmap towards a preferred future outcome. In this context, Guston defines anticipatory governance as 'a broad-based capacity … that can act on a variety of inputs to manage emerging knowledge-based technologies while such management is still possible.'[55] Planning into the long-term horizon involves some uncertainty but offers flexibility in designing a regulatory policy response.[56]

The insights generated may inform policies including early action to invest in funding technological development efforts. A particular technology could be regarded as holding real promise even while at a low technology readiness level. The strategic decision may therefore be made to provide support on its journey from 'lab-to-fab'. In making decision on what to fund, the European Innovation Council, whose objective is to ensure that the European Union is at the fore of deep tech innovation, uses 'an anticipatory lens' and 'looks to identify technologies emerging from the science base that could create new value propositions and/or disrupt existing models.'[57] At this point strategic foresight is an integral part of how the European Commission works in order to develop future-proof policies including through having regard to 'emerging megatrends in the green, digital, geopolitical and socio-economic contexts.'[58] The EU Policy Lab engages in foresight and horizon scanning, both to anticipate emerging challenges, and to consider how policy choices may work out in the longer term.

One of the most striking examples of foresight activities driving anticipatory governance has been the EU's determined policy drive to adopt legislation governing the deployment of AI in the EU. Back in 2017, the European Council called for 'a sense of urgency' to be galvanised to addressing emerging trends such as AI.[59] What resulted became the European Commission's landmark EU AI Act Proposal[60] with a view to ensuring responsible innovation and protecting fundamental rights. The AI Act was designed to be robust but flexible while being 'comprehensive and future-proof in its fundamental regulatory choices.'[61] A further objective was the facilitation of the development of a single market for trustworthy AI. While introducing a hierarchical risk-based system, it also involved leaving room for best practice industry-led standards to evolve. The European Commission's groundbreaking work on this began when the territory and how it would evolve was far from fully mapped and during the legislative process, the arrival of Generative AI presented novel challenges that needed to be weighed up. The reaching of agreement of the EU AI Act delivering a harmonised, risk-based rulebook on AI was a real feather in the regulatory cap of the European Union.[62] Furthermore, fittingly within the spirit of anticipatory governance, the European Commission aims to counteract the transition time lag to the AI Act becoming operational by encouraging companies to sign up to a voluntary AI Pact.[63]

To achieve maximum benefit, horizon-scanning and future-gazing should be continuously undertaken and their insights embedded into policy design and to ensure resilience. For this to work effectively, a variety of technical expertise as well as good stakeholder relationships and innovative implementation methods are needed. This goes beyond simple trendspotting. Analysis needs to be done to identify emerging trends and what they mean for our society and policy development in the short, medium and long-term. Sense-making and the dissemination of the policy significance is crucial. There are green shoots in this regard. For example, the European Commission publishes an annual Strategic Foresight Report to inform the development of its work programmes.[64] Its content is developed on the basis of participative and cross-sectoral foresight processes. In the United Kingdom, the Emerging Technology ('EmTech') Research Hub at the Financial Conduct Authority conducts research to identify critical and emerging technology trends for financial services in the medium and long-term to inform regulatory thinking across policy and strategy in the area of financial services.[65] EmTech is exploring synthetic data and privacy enhancing technologies, DLT, Quantum Technologies, and the Metaverse and immersive technologies.[66] Insights are gathered using research papers, responses to calls for input, working groups and workshops. It also links in with the Digital Regulation Cooperation Forum. Activities such as these demonstrate that regulators are beginning to embody the values of strategic regulatory foresight as a central part of their working methods.

---

[54] See further Government Office for Science (UK) (2017).
[55] Guston (2014), p. 219.
[56] Soeteman-Hernández et al. (2021).
[57] European Innovation Council (2023), p. 7.
[58] European Commission (2023b).
[59] European Council (2020) Special meeting of the European Council, 1 and 2 October 2020 – Conclusions.
[60] European Commission (2021).
[61] European Commission Proposal, Explanatory Memorandum , 1.1.
[62] Of course evaluation awaits the final text and the experience of the Regulation's roll-out on the ground.
[63] European Commission (2023a).

## 5  Iterative Policymaking

As elaborated upon in Section 2, traditional incremental approaches to regulating[67] that rely on lengthy procedures are simply no match for disruptive and rapidly evolving innovation. Even regulatory impact assessments, a hallmark of the better regulation agenda, come from a 'one-shot' policymaking mindset[68] that falls short -- focusing on a present view rather than a far-reaching, future-based one. An anticipatory governance regulatory culture has as its lodestone a genuine willingness to deliver disruptive but timely policy change where warranted though bold and iterative policymaking over time. At an advanced level this type of regulatory ecology should have mechanisms that are removed from the outmoded nature and process of traditional regulation to produce a far more reflexive and dynamic model. Iterative policy making, rather than being one-off, allows more than one chance at evolving policy iteration, learning from what went before. Post-implementation reviews are vital to ensure that regulation stands up and does not become outdated.[69] The values of regulatory experimentation and regulatory learning are present. While iterative policy development has been used across jurisdictions in a number of fields including health and social care, it remains a concept under development.

Scholars such as Braithwaite[70] and Sparrow[71] have emphasised the importance of regulatory tools being devised in response to problems rather than the response to problems being driven by a limited pre-existing menu of regulatory approaches that can bake in blind spots. This resonates here. Characteristically, within anticipatory governance a pre-determined approach or outcome should be eschewed in the pursuit of iterative policymaking. This can yield more thoughtful regulatory solutions. With iterative policymaking as part of an anticipatory governance agenda, regulatory agencies can test regulatory ideas to see if they produce the expected results or need to be tweaked to better ensure that they meet their regulatory goals. This is by nature a collaborative process. Iterative regulation for innovation policy is often quintessentially a process of co-creation bringing together a diverse set of actors and expertise to facilitate trial and error. Such stakeholders may include regulators, industry, civil society, environmental watchdogs, academia and non-governmental agencies ('NGOs'). While stakeholder consultation has been an entrenched feature of policy development for some time, these approaches amount to a more equal status of co-creation. There is also respect for transdisciplinary methods in the task of co-analysis and co-creation. The aim should not be to be visionary but to experiment freely and cooperatively and see what sticks. Exploring multiple futures allows for a focus on what should be done to anticipate these scenarios if they seem likely.

---

[64] European Commission (2023f).
[65] Financial Conduct Authority (2023b).
[66] Financial Conduct Authority (2023b).
[67] Kingdon (2003), pp 79-83.
[68] Baldwin (2010), p 265.
[69] Regulatory Horizons Council (2022), p. 7.
[70] Braithwaite (2005); Braithwaite (2008).
[71] Sparrow (2000).

An iterative approach to policymaking is inherently flexible and often involves a role-shift, putting the State in the form of co-creator rather than gatekeeper.[72] Broad, stakeholder-informed approaches can be seen in the Digital Charter developed in Canada. When passed, the Digital Charter Implementation Act, 2022 will legislatively implement the Charter which provides a privacy protection framework for data-driven innovation in the form of 10 principles. Such agile, adaptive policy development through continued improvement has the potential to come into its own for emerging technologies and applications as well as entirely new technologies such as quantum technologies.

Iterative policy and regulatory development of principles and regulation concentrate on getting regulatory policy design off the ground at an early stage with a view to offsetting the delay that typically comes with prolonged study of the effects of innovation on markets and consumers. This is seen in the regulatory competition stroke pulled off by the European Union in choosing to move quickly to regulate, even in an imperfect manner, established aspects of crypto-assets in MiCA.[73] The decision was made to leave more complex and evolving pieces of the jigsaw for a later day.[74] This is the epitome of timely, iterative anticipatory governance. While enabling innovation may be a central goal, the application of the precautionary principle should not be forgotten when engaging with emerging innovation.[75] However, the risk of getting it wrong or missing an angle is offset by multi-stakeholder involvement and the capacity to improve the next iteration. Furthermore, sophisticated advance modelling can be deployed to reduce the risk of getting it wrong. The European Union's Joint Research Centre plans to use a pre-mortem approach, deploying a series of challenging scenarios against which to stress test potential regulatory frameworks. This will provide invaluable upfront feedback on the anticipated performance of a draft formulation of a policy instrument against regulatory goals.[76] Equality and diversity perspectives also need to be fully integrated into policymaking for innovation including risks of technological and financial exclusion.[77] Technological exclusion can arise where sections of the population that do not have high digital literacy are given no option but to use technology to access a good or service.[78] Financial exclusion can arise particularly for older sections of the population who are not digitally literate and/or prefer to do their business such as banking in person rather than through digital banking. This is the classic digital divide – digital advancement and digital services do not yield uniform outcomes for all sectors of society and business. As the OECD acknowledges, to achieve an inclusive digital economy requires careful planning.[79] To address the digital divide, we need to consider how to support segments of the population that are digitally vulnerable in 'the digital-by default world'.[80] Diverse representation and stakeholder involvement helps to ensure that discussion and outcomes are balanced through considering the opportunities and risk around innovation from a wide variety of perspectives within the lens of the wider population including how issues such as age, disability, race and income may lead to differential impacts.

---

[72] Soeteman-Hernández et al. (2021), p. 100301.
[73] Markets in Crypto-Assets Regulation (EU) 2023/1114.
[74] Maume (2023).
[75] Rübig (2012).
[76] Joint Research Council (2022), p. 22.
[77] Selwyn (2002).
[78] Seifert (2020).
[79] OECD (2021).
[80] Heuer et al. (2022), p. 27.

An example of iterative policymaking (as opposed to more traditional incremental policymaking) is provided by how the United Kingdom has approached governing AI. The objective is to create cross-sectoral governance principles in relation to the opportunities and risks associated with AI. Interestingly, once devised, these principles will at first be available on a non-statutory basis for use by regulators at their discretion.[81] This will provide the opportunity to adjust course if the principles are not judged effective in encouraging innovation while suitably mitigating risk. The possibility is also retained for the principles to be subsequently made the subject of a statutory duty on regulators to have due regard to the principles.[82] This fresh approach brings a regulatory culture that actively moves the needle by providing guiding principles for regulators on AI but still gives space for experiential learning to inform future iterations of the policy result rather than immediately adopting legislative principles that may prove ill-adapted but ossified. The iterative approach taken here to rolling-out cross-sectoral governance principles for AI demonstrates the essence of an anticipatory regulatory culture, allowing room to course adjust. However, while this flexibility and discretion is helpful to allow leeway for context, unless carefully handled, it may cause uncertainty for regulators in relation to how they should do their job. More optimistically, perhaps this type of wobbling should be viewed as a necessary part of an anticipatory regulatory culture as it totters towards the future. The lesson of an anticipatory regulatory culture is that, much like innovation, we need to be able to embrace uncertainty, not run from it if we are not in a position to eliminate it. However, this is not a case of 'one and done'. Eventually a firmer regulatory approach may be called for.

## 6    Regulatory Experimentation Mechanisms

Regulatory experimentation is an obvious facet of the new anticipatory governance. However, the concept suffers from a lack of full conceptualisation. Experimentation can involve 'taking action to ensure that … policy approaches work.'[83] The intention of such experimentation is to ensure that the outcomes of policy approaches are those intended and to consider unintended consequences and if necessary go back to the drawing board. Regulatory experimentation that is focused on regulatory compliance or regulatory change ought to be distinguished from pro-innovation policy drivers whereby regulators initiate mission-based challenges that challenge innovators to devise solutions to pressing challenges for our society and economy. These may create a temporary model regulatory framework incentivising technological rather than regulatory innovation. However, mission-focused and regulatory learning objectives are not incompatible and regulatory learning can be an indirect result. The term 'experimental regulation' also lacks an agreed definition.[84] Here the term 'regulatory experimentation' is preferred over 'experimental regulation' as it is broader in its embrace of not just the trialling of regulation, but all forms of experimentation associated with delivering the building blocks of anticipatory governance.[85] As defined by the Canada's Centre for Regulatory Innovation, '[a] regulatory experiment is a trial or test of a new product, service, approach or process designed to generate evidence or information that can inform the design or administration of a regulatory regime.'[86]

Regulatory experimentation can be used by regulators to make complex regulatory policy decisions where uncertainty exists as to how a contemplated regulation will apply because the field is cutting-edge or evolving rather than settled. It is the norm for pharmaceutical companies to engage in clinical trials to test new drugs to see if they achieve the intended aims and it is eminently sensible for regulatory agencies to take the same approach in relation to innovation. Regulatory experimentation can likewise be used to support regulatory decision-making by providing legislators/regulators with real-world evidence to underpin decisions as to whether to proceed as planned or to make adjustments or go back to the drawing board. Testing regulation can supply information on the impacts and potential application in practice and how it could affect industry and society and whether the appropriate balance between enabling and deterrence has been achieved.

---

[81] Secretary of State for Science, Innovation and Technology (UK) (2023), paras 55-59; Department for Science, Innovation and Technology (2024), para 16.
[82] Secretary of State for Science, Innovation and Technology (UK) (2023), paras 55-59.
[83] Tõnurist and Hanson (2020), p. 2.
[84] Ranchordás (2021), p. 4.
[85] Regulatory sandboxes, pilot regulation, policy labs and experimentation clauses are diverse forms of experimentation are delved into below before we turn to the subject of regulatory learning.
[86] Centre for Regulatory Innovation Canada (2023).

Experimentation mechanisms go beyond how think tanks function as 'talking shops', holding the potential to radically 'bridge the gap between what we know and what we do'.[87] However, successful regulatory experimentation is not achieved without careful attention to design choices. This should also involve multi-stakeholder 'participatory foresight'.[88] Within the rubric of regulatory experimentation, experimental approaches to policymaking for innovation reflect a call within policy science for policymaking to draw from product design and scientific invention.[89] A key element of this regulatory experimentation involves bringing product design conceptualisation to bear in developing policy solutions to the issues identified.[90] At the most advanced level, creativity and storytelling are used to plausibly speculate about the future developments in terms of opportunities to be enabled and potential risks to be curbed.[91] The EU Policy Lab has mastered this at a high level. Its policy labs have involved co-creation of fictional artefacts designed to spark 'forward-looking discussions into the possibility or plausibility of yet to be fully fledged realities.'[92]

### 6.1  Regulatory Sandboxes

Regulatory sandboxes have proved to be one of the most emulated styles of experimentation or controlled testing under regulatory supervision. A key benefit is that regulatory sandboxes offer two-way learning opportunities for tech innovators and regulators as innovators trial their products and seek to navigate an ill-fitting regulatory framework not specifically designed with the innovation in mind, and regulators learn about technical and business model goals. Within an anticipatory governance lens, this type of concentrated regulatory learning boost contributes immeasurably to both regulatory and technical capacity building. A clear benefit for regulators operating regulatory sandboxes lies in being able to identify regulatory obstacles.

Initially pioneered by the UK's Financial Conduct Authority,[93] regulatory sandboxes have proliferated globally[94] and within Member States of the European Union[95] as a means of navigating regulatory uncertainty around the roll-out of technological innovation. Having initially stood back as regulatory sandboxes developed, the European Union is now eagerly embracing their versatility in a range of public and private contexts with a view not just to reducing regulatory uncertainty but also to stimulating innovation. The Interoperable Europe Act[96] is designed to enable interoperable regulatory sandboxes to be established by the European Commission for public sector testing of innovative solutions designed to achieve cross-border digital interoperability in the provision of public sector services. The aim is that such sandboxes 'should not only contribute to new technological solutions but also to regulatory learning'.[97] Meanwhile the European Blockchain Regulatory Sandbox established in 2023 is enabling cross-border regulator-innovator dialogue around a pan-European framework for experimentation with blockchain solutions for the private and the public sector.[98]

---

[87] Lee and Ling (2020), p. 285.
[88] Hebinck et al. (2018); Muiderman et al. (2020); Fenwick and Vermeulen (2020.
[89] Parsons (2002).
[90] See further Mintrom and Luetjens (2016).
[91] See Strengers et al. (2022).
[92] Pólova and Nascimento (2021), p. 7.
[93] It has also developed a Digital Sandbox for firms to use data sets to test and build prototype models. In addition, the United Kingdom's Information Commissioner's Office has a Regulatory Sandbox for products and services using personal data.
[94] Allen (2019); Ahern (2019); Miglionico (2022).
[95] Ahern (2021).
[96] European Commission (2022a).

The European Union has also come to see the value of regulatory sandboxes in contributing to the competitiveness of European business including small and medium-sized enterprises ('SMEs') by helping to tackle regulatory ambiguity head-on. The Commission's 2023 SME Relief Package recognises the place for sandboxes to enable experimentation in a controlled yet real-world environment where regulatory opacity or complexity exists that is resource-intensive to navigate.[99] There is thus a commitment to taking action to work alongside Member States to promote regulatory sandboxes for experimentation to drive innovation by start-ups. From an anticipatory governance perspective, over time these regulatory sandboxes may provide the requisite boost to regulatory learning to deliver the stimulus for 'a more innovation-friendly and future-proof regulatory framework'.[100]

Arguably the most notable development thus far for regulatory sandboxes at EU level is the significant role that they are expected to play in the EU AI Act.[101] AI regulatory sandboxes in this context are expected to deliver the twin objectives of (i) facilitating safe and controlled testing under regulatory supervision, and (ii) assisting companies to comply with the regulatory burden associated with compliance with the AI Act and navigating impacts on fundamental rights. This is a tall order for competent authorities. As matters stand there are many complexities and unknowns as to how sandboxes under the AI Act will operate in practice[102] but the availability of cross-border, inter-jurisdictional sandboxes should help in terms of cooperation on resources and capacity-building as well as consistency of approach. The pilot AI Regulatory sandbox that is backed by the European Commission is being counted upon to deliver insights on lessons learned and best practices that will be formally reported upon to the European Commission to assist with developing EU conformity guidelines.[103] Regulatory learning will also come from observing Spain's experience as the first Member State to legislate pre-emptively for AI regulatory sandboxes that will operate under the AI Act.[104] Spain's agile legislation is designed to allow companies and the public sector to engage in testing of their AI systems to get their compliance efforts off the ground well in advance of the AI Act coming into effect.

### 6.2 Pilot Regulation

Pilot regulation concerns experimenting through trial regulation, testing the effectiveness of a proposed regulatory framework in meeting its regulatory goals by means of temporary scaled-down testing. As such, it fits hand in glove into an anticipatory governance agenda. Pilot regulation can come into its own in finding the sweet spot between over-regulation and under-regulation of innovation. Top-down pilot regulation projects have emerged as a relatively new form of regulatory experimentation. Pilot regulation projects are typically initiated by a government agency or regulator in order to engage in a real-world evaluation of the suitability of the proposed form of regulation for innovation.[105] They can include both entirely new regulation and/or proposed modifications to existing regulation to tailor it for innovation. As the European Commission observes, pilot regulation 'reflects a clear predefined focus field for experimentation by the regulator, with an identified need to change some aspects of the existing regulation. Even if the direction is clear, the concept has to be tested before any permanent change is introduced.'[106] The hallmark of pilot regulation is proactiveness from a regulator who is already familiar with the policy dilemmas and the regulatory gaps but who wants evidence from the field to buttress and future-proof the regulatory policy decisions taken. Pilot projects can be legislative or devised with minimum formality as compared with the lengthy, drawn-out processes of legislation. Rather than participation by market actors being by way of invitation, ideally it should be open to relevant market participants to opt in to the trialling of pilot regulation based on an open call. This is consistent with the European Commission's description in its Staff Working Document on experimental regulation which provides the following description:

> 'Pilot regulation consists of a temporary regulatory framework applicable on a voluntary basis. This may be applicable to all market actors or a group of them (e.g. distribution system operators). There is no selection procedure, but it is open to all or a specific group of market actors for voluntary participation and the pilot is therefore non-discriminatory.' [107]

---

[97] European Commission (2022a), p. 12.
[98] https://ec.europa.eu/digital-building-blocks/sites/display/EBSI/Sandbox+Project. Accessed 12 March 2024. See also European Commission (2024).
[99] European Commission (2023e).
[100] European Commission (2023e), p. 6.
[101] European Commission (2021).
[102] Buocz, Pfotenhauer and Eisenberger (2023).
[103] European Commission (2022c).
[104] Spanish Royal Decree 817/2023 https://www.boe.es/boe/dias/2023/11/09/pdfs/BOE-A-2023-22767.pdf. Accessed 12 March 2024.

Market participation in the experiment assists the agency leading the project in assessing regulatory options as they play out in the ground before making a choice on which regulatory choice is optimal. The objective of the pilot regulation trial is for regulatory learning to occur from direct observation and feedback which will inform decision-making on the final form of the regulatory framework.[108] One example of such approach comes from Canada. Transport Canada has tested light sport aircraft to see whether it could be used in a pilot training environment. This required an experiment with the participation of flight schools involving regulatory flexibility to allow what would otherwise be a prohibited practice. Data gathered regarding aircraft reliability, emissions and noise was used as regulatory learning to inform a decision on whether permanent regulatory changes were warranted.[109]

Sometimes pilot regulation is used with the primary goal of promoting technological innovation, with regulatory innovation being a by-product. A prime example is provided by the EU's Distributed Ledger Technology ('DLT') Pilot Regulation ('DLTPR'), [110] a deliverable of the European Commission's Digital Finance Package.[111] The DLTPR created a temporary, opt-in pilot regime for DLT-based market infrastructure (most relevantly using Corda, Ethereum and Hyperledger Fabric) for trading and settlement of tokenised financial instruments. The three year pilot regime which began in March 2023 permits the issuance[112] and transfer of tokenised assets through the use of DLT and settlement using tokenised money. The objective is to allow DLT Pilot participants to 'flexibly experiment with DLT in organizing trading and settlement of financial instruments'.[113] An anticipatory governance motivation is evident in the Recitals to the DLTPR:

> 'It is important to ensure that Union financial services legislation is fit for the digital age and contributes to a future-proof economy that works for citizens, including by enabling the use of innovative technologies. The Union has a policy interest in exploring, developing and promoting the uptake of transformative technologies in the financial sector, including the uptake of distributed ledger technology (DLT).'[114]

---

[105] A variation occurs where a suitable regulatory framework has not yet been devised by the legislator/regulator for an identified scenario and the goal of the project is to assist in proactively devising it. This is the primary focus of a policy lab.
[106] European Commission (2023d), p. 53.
[107] European Commission (2023d), p. 12.
[108] Pilot regulation schemes are sometimes accompanied by funding support but this is not always the case.
[109] Government of Canada, What is regulatory experimentation? https://www.canada.ca/en/government/system/laws/developing-improving-federal-regulations/modernizing-regulations/regulatory-experimentation.html. Accessed 12 March 2024.
[110] DLT Pilot Regulation (EU) 2022/858.
[111] European Commission (2020).
[112] Issuance is by digital representation of a financing instrument on DLT or by issuing what began life as a traditional financial instrument in tokenised form as a DLT financial instrument.

The DLTPR contains experimentation clauses which provide a framework for granting regulatory exemptions. The granting of temporary regulatory exemptions is designed to encourage experimentation that might not otherwise happen.[115] Market participants must apply for permission to operate a DLT Market Infrastructure under the DLT Pilot Regulation and expressly justify each request for a regulatory exemption. In the case of a DLT multilateral trading facility, applicants must demonstrate that each exemption is proportionate to and justified by the use of DLT. The potential for regulatory learning is assisted by the fact that the DLTPR permits retail investors to have direct access to DLT market infrastructures and related financial instruments (although the nature of such transactions is restricted to ensure investor protection).

In due course the European Securities and Markets Authority ('ESMA') must present a review report to the European Commission on DLT market infrastructures. This will cover matters such as technical issues, as well as potential risks to investor protection, market integrity of financial stability. The report must also make an overall cost-benefit assessment leading to a recommendation as to whether the DLTPR should be continued and under what conditions. The fervent hope is that if the pilot regime proves a success, it could be put on a permanent footing through legislative amendments to EU financial services legislation 'to establish a single coherent framework.'[116] As a form of pilot regulation, the DLTPR is thus an important pioneer in the European Union's fledgling adoption of elements of an anticipatory governance culture through encouraging the application of DLT in new contexts and providing a regulatory container for it.

### 6.3 Policy Labs

The promise of policy labs within an anticipatory governance culture is huge. However, a difficulty for the study of policy labs is that 'conceptual indeterminacy'[117] continues to present a challenge. The definition of 'policy ' adopted here is a time-bounded and issue-bounded mechanism established by policymakers and/or regulators that engages stakeholders in studying a regulatory or policy challenge with a view to assisting in the task of devising regulation or policy for innovative technologies in application.[118] To this understanding we can add Nesta's description of a policy lab as 'a group of actors with different skills aiming to develop regulation. The policy lab uses a set of user-centred methods and competencies to test, experiment and learn in policy development.'[119]

---

[113] European Securities and Markets Authority (2023), p. 18.
[114] DLT Pilot Regulation (EU) 2022/858, Recital 1.
[115] The DLT Pilot Regime also responded to regulatory lacunae surrounding the use of distributed ledger technology and the direct platform-based trading of crypto-assets that meet the definition of financial instruments, as well as the potential for novel risks to arise from the technology.
[116] DLT Pilot Regulation (EU) 2022/858, Recital 53.
[117] McGann et al. (2018).
[118] For example, the Policy Learning Labs organised by the Krakow Living Lab https://www.kpt.krakow.pl/en/laboratories/livinglab/. Accessed 12 March 2024.
[119] Arntzen et al. (2019), p. 79.

Alternative terms to 'policy lab' are in use the public domain, for example, 'public sector innovation labs', 'government innovation labs', 'living labs', 'public innovation labs' and 'social innovation labs'. In addition, the mantle 'policy learning lab' is sometimes used to describe what amount simply to policy innovation workshops and briefing activities for policymakers.[120] For present purposes we focus on state-sponsored policy labs that feed into public policy development as opposed to labs that are focused on delivering mission-focused social or community innovation with no underlying public policy formation driver. Ownership of policy labs may arise in a number of ways: (i) governmental or government department facilitated (even if located tangentially)[121] (ii) researcher/university facilitated[122] (iii) private sector facilitated or (iv) a hybrid form. Policy labs that are state-driven are driven by the public sector even if the relevant agency has a semi-autonomous status or a hybrid form. Where ownership arises at governmental level, a separate and independent structure can allow for inventive and autonomous working methods that also bypass policymaking bottle necks.

The role of policy labs as agents of change has been highlighted in the literature.[123] However, despite considerable evidence of the use of policy learning labs in the sense defined here,[124] the phenomenon remains underexplored at a scholarly level. With some exceptions,[125] much of what is written contemporaneously falls into the category of guides to the labs themselves or practitioner-generated work, reflecting the fact that the field is still emerging. Policy labs can generate regulatory learning from innovators around regulatory barriers encountered including useful perspectives garnered from private citizens and civil society actors. Creative experimentation in policy labs can provide the bridge to prescient policy ideas, for example, to envision how cutting-edge technologies could be used to accelerate the green transition of the energy system through next generation smart grids and digitalisation of energy systems. However, to date, the mandate of policy labs has often focused on achieving societal policy objectives rather than on making specific gains in innovation regulatory policy. Nonetheless, policy labs hold real potential to instrumentally serve anticipatory governance goals. Complex, knotty issues that raise a plethora of challenges lend themselves to 'labbing'. Policy labs can take advantage of timing when a policy window emerges within which to tackle an important but challenging regulatory and/or policy lacuna that has been identified in regulatory foresight work. A similar initiative is a policy sprint, which can accelerate policy development by bringing together stakeholders to agree on directions and potential solutions.

Although they differ in terms of organiser and funder, the essential characteristics of policy labs are commonly found across all types whether public or private. First, policy labs offer a neutral forum where participants from academia, government, public sector, business and citizens can come together with a view to reaching a common policy goal. Second, a policy lab provides an engagement mechanism that facilitates the incorporation of evidence into policy.[126] Providing evidence to policymakers is one thing; they must be open to considering its impact on their decision-making rather than coming with pre-determined approaches which are rigidly adhered to. The hallmark of a successfully designed and operated policy lab is the openness of the participants to discussion around a well-identified issue on which they have been well briefed at a level tailored to their likely level of expertise and understanding.

Some examples of policy labs' focus can be usefully summarised for illustrative purposes. Policy labs have been used in Sweden as part of the Swedish Testbed Strategy to tackle complex challenges round innovation such as how to integrate autonomous vehicles into society.[127] In the United States the Public Policy Lab[128] works on projects that intersect policy areas and multiple regulatory agencies with a view to making systemic change. One of these projects with the National Science Foundation is considering the design and piloting of concepts for how US federal seed-funding programmes could increase engagement and participation from entrepreneurs from underrepresented communities. SINTEG was a German initiative[129] to identify, not just model energy solutions for the future, but also how to adapt the regulatory framework to serve a smart energy world so that the regulatory environment would be conducive to rolling out solutions that would be renewable and underpinned by digital technology.[130] In Sweden, Research Institutes of Sweden (RISE) hosts a Policy Lab for policy and regulatory development. They note that for technological development to reach its full potential, policy and regulatory changes must support it. Recognising that engineers, computer scientists, lawyers, economists and others must work collaboratively, RISE coordinates the bringing together of the requisite competencies and skills.

---

[120] For a guide to this see Bailey et al. (2019).
[121] For example, the UK Policy Lab https://openpolicy.blog.gov.uk/about/. Accessed 12 March 2024.
[122] Some of these policy labs have been funded or co-funded by the European Union.
[123] Schuurman and Tõnurist (2017).
[124] Gofen and Golan (2020).
[125] McGann et al. 2018).
[126] Hinrichs-Krapels et al. (2020), p. 1.

The EU Policy Lab established in 2014 enables cross-disciplinary involvement in policymaking, leveraging insights from the Joint Research Centre in the European Commission. Integral to its work is a stakeholder collaboration approach that makes use of design and visual communication tools in the co-creation of innovative policy protypes. Its latest project involves partnering with national policy labs around drawing attention to water resilience issues at EU level.[131] It provides 'a physical space designed to foster creativity and engagement to develop interactions, processes and tools contributing to bring innovation in the European policy-making.'[132] The EU Policy Lab describes itself as 'a mindset and a way of working together that combines stories and data, anticipation and analysis, imagination and action.'[133] First and foremost, policy labs like this offer bespoke opportunities for a diversity of stakeholders to come together. Second, key to any policy lab is the role of the facilitator who brings the parties together. This is in line with the concept of 'knowledge brokerage for policymaking'[134] where the scientific community helps to ensure that whatever policy emerges is informed by evidence but avoids acting as a direct driver of policy choice. The EU's Joint Research Centre ('JRC') recognises this role of honest broker.[135] In line with this, leaders of policy labs who design and facilitate them know that building a dynamic of trust and openness is key to facilitating shared iterative decision-making.

Policy labs work by (i) providing a forum for open dialogue around a policy issue; (ii) allowing the formation of networks and collaborations between researchers and policymakers; and (iii) synthesising a robust evidence base for the policy topic gathered in order to catalyse development and testing of an appropriate policy response.[136] In order for policy labs to work, careful planning is needed to correctly frame the policy problem and the steps that will be undertaken to interrogate it and to what end. Indeed, seasoned operators suggest that around 20-40 staff days go into the design of each topic-specific policy lab.[137] This involves, for example, getting a full picture of regulatory actors, policy and regulation in the area and identifying the regulatory gaps. It is also important to think carefully about who to invite to participate and to understand their background and level of expertise.

---

[127] Arntzen et al. (2019).
[128] https://www.publicpolicylab.org/projects/. Accessed 12 March 2024.
[129] Although labelled as a regulatory sandbox approach SINTEG's characteristics more closely fit those of a policy lab.
[130] https://www.bmwk.de/Redaktion/DE/Dossier/sinteg. Accessed 12 March 2024.See also Kert et al. 2022, p. 5.
[131] On such labs for water see further Ojha, Maheshwari and Bhattarai (2021).
[132] Policy Lab https://joint-research-centre.ec.europa.eu/laboratories-and-facilities/policy-lab_en. Accessed 12 March 2024.
[133] 'We are the EU Policy Lab' https://policy-lab.ec.europa.eu/index_en. Accessed 12 March 2024.
[134] Gluckman et al. (2021), p. 3.
[135] Topp et al. (2018).
[136] See further Hinrichs-Krapels et al. (2020).
[137] Hinrichs-Krapels et al. (2020), p. 3.

Communications and documents must be pitched appropriately so that non-technical people and non-specialists can grasp the issues and their policy significance.

Testing and probing are part of the modus operandi of policy labs. Policy labs also straddle regulatory foresight activities and can focus on 'the practical bridge between foresight and design fictions'[138] where experiential experimentation takes places around possible futures. In this vein, the European Commission's JRC Living Labs can combine testing and regulatory learning.[139] An EU Policy Lab based in Brussels being created for the Horizon Europe PLAN'EAT project involves EU and national policymakers working to provide an interface for science and food policy within a broader project involving a network of Living Labs.[140] The associated policy lab for this project will use insights gained to test and develop systemic policy solutions. The United Kingdom has gained relevant experience through an established history of open policymaking through the lens of a civil service Policy Lab.[141] It involves cross-government inter-disciplinarity in its methods which 'are grounded in evidence, participation and experimentation' to both 'tackle complexity and build complexity.'[142] One of its non-traditional approaches to devising policymaking is to involve users in co-creation, testing and learning.

The tangible contribution that individual policy labs make to regulatory learning will depend on each particular policy lab's design, its leadership, not to mention the level of political will to embrace its outputs. This is why policy labs ideally need to be a strategic part of a joined up anticipatory governance culture. Some approaches identified in policy labs to how to address a particular policy problem may provoke controversy among lab participants and these will have to be evaluated and skilfully handled by the policy lab team in order to build a robust consensus where possible as to what is desirable and feasible in policy terms. Policy labs do vary in terms of the ambition and targeting of their outputs which could include a policy briefing document or report. The difficulty with the inherent nature of policy lab is that there is no guarantee that any policy implications teased out and solutions identified will actually be addressed once the policy lab draws to a close and its findings are disseminated to participants. This depends in part on political goodwill, resourcing realities and governmental priorities. Nonetheless, the seed may be sown for further work on policy development in the area based on the evidence presented and the ideas generated by participants in the course of a collaborative policy lab between researchers and policymakers.

## 6.4 Experimentation Clauses

Experimentation clauses are another component of a regulatory toolbox for innovation. Whether provided for in legislation or other regulatory frameworks, experimentation clauses provide a mechanism for competent authorities or regulators to provide regulatory flexibility to innovators during defined experimentation. The European Council's Conclusions on Regulatory Sandboxes and Experimentation define experimentation clauses as 'legal provisions which enable the authorities tasked with implementing and enforcing the legislation to exercise on a case-by-case basis a degree of flexibility in relation to testing innovative technologies, products, services or approaches.'[143] An experimentation clause is an especially useful mechanism where the competent authority or regulator is not already vested with discretionary power that can be exercised to allow regulatory flexibility for innovators to experiment or test. An experimentation clause is equally useful where regulation would otherwise not afford any margin of appreciation. It is common to specify a maximum duration for an experimentation clause to apply (including, where applicable, extensions to an original derogation period granted). Regulatory insights gained through experiential application of a specific derogation pursuant to an experimentation clause during testing may well form the seeds for updating the law to more appropriately meet future needs arising from the scenario tested for.[144] Alternatively, as with pilot regulation, where merited, the type of derogation exceptionally provided for through an experimentation clause could ultimately be legislatively established on a more regular footing where it proves itself to be apt to continue beyond its status as a stopgap measure.

---

[138] Pólova and Nascimento (2021), p. 12.
[139] See Alonso Raposo et al. (2021) and the JRC Living Lab for Digital Energy Solutions https://www.eranet-smartenergysystems.eu/ll/143/JRC-Living-Lab-for-testing-digital-energy-solutions.html. Accessed 12 March 2024.
[140] https://planeat-project.eu/policy-lab-brussels/. Accessed 12 March 2024.
[141] https://openpolicy.blog.gov.uk/about/. Accessed 12 March 2024.
[142] https://openpolicy.blog.gov.uk/about/. Accessed 12 March 2024.
[143] European Council (2020).

Experimentation clauses provide regulatory flexibility that can help to drive innovation. As a pro-innovation tool, experimentation clauses target the effects of regulatory lag while promoting the uptake of technology to tackle major societal challenges. Looked at in terms of regulatory competition, a jurisdiction's willingness to integrate experimentation clauses within its innovation strategy and regulatory framework may enhance the competitiveness of its offering in the eyes of innovators. Indeed, some EU Member States have warmly embraced experimentation clauses as part of an overall innovation strategy. France's adaptive 'France Expérimentation' regime is an exemplar in this regard. Billed as removing regulatory (but not legislative) blockages in the way of ambitious innovators' projects (including projects with purely economic ends and projects designed to accelerate climate transition goals), France Expérimentation provides a single point of contact from which communication is opened with all relevant ministries and regulatory bodies.[145] The scheme was used for ROBAGRI, a project enabling agricultural robots to circulate on certain public roads. In Germany, there is a clear mandate to use experimental clauses to propel innovation and they have been used in relation to passenger transport law for testing ride-sharing and automated buses. The Federal Ministry for Economic Affairs and Energy (BMWi) has shown leadership in issuing guidelines on the formulation of experimentation clauses.[146]

At an EU level, experimentation clauses already form part of the EU acquis[147] and there is an appetite to expand this regulatory approach. An example is provided by the Industrial Emissions Directive[148] which is the primary EU instrument regulating industrial emissions. Competent authorities can grant temporary derogations for up to 9 months from certain emission level requirements.[149] The Commission's Proposal for a revised Industrial Emission Directive would allow operators testing emerging techniques to enhance environmental performance the opportunity to avail of expanded regulatory derogations for up to 24 months.[150] Given burgeoning interest in anticipatory governance, it seems entirely possible that experimentation clauses will also bear fruit in other EU legislative contexts. That said, attention to proper design is critical.

For experimentation clauses to function optimally clarity in purpose and design is crucial. To work properly, experimentation clauses need to be precise and subject-specific, delineated to apply to a well-defined regulatory scope (both operational and legal), rather than purporting to be all-purpose, which apart from suffering from uncertainty, would have the potential to undermine regulation. If we take the example of a self-driving vehicle, the real-world testing being undertaken and its purpose need to be clear (e.g., controlled testing of autonomous vehicles to see how they perform on a real road), as well as the regulatory rules that are open to derogation (for example, certain rules requiring vehicles to be driven by licensed drivers).

---

[144] On this see France Expérimentation (2023a).
[145] France Expérimentation (2023b).
[146] Federal Ministry for Economic Affairs and Energy (Germany) (2020).
[147] See, for example, Article 4(5) of the Weights and Dimensions Directive 96/53/EC which enables Member States to authorise vehicle use for a trial period during which the Directive's requirements do not have been complied with.
[148] Industrial Emissions Directive 2010/75/EU (Recast).
[149] Industrial Emissions Directive 2010/75/EU (Recast), Article 15(5).
[150] European Commission (2022b), Article 27b.

When providing for experimentation clauses that dilute the impact of regulation, regard ought also to be had to the potential risks attendant on derogating from legal requirements that would otherwise be considered apt. Risk is inherent in testing emerging technologies and new applications and business models. Consequently responsibility and caution are needed. This is aided by providing discretion to the authorities as to when is appropriate to provide regulatory flexibility in a given case and when it is not. Consistent with this, experimentation clauses for novel technological testing purposes should be expressed as permissive rather than mandatory in nature, allowing a case by case approach by the competent authority. The authorities can also make use of conditional provisos that must be satisfied with a view to safeguarding public interest considerations. Other risk-minimisation preconditions for experimentation could be imposed relating to the qualifications and experience of relevant personnel, insurance cover and expert assessments. It is also prudent to provide a broad power to suspend or revoke an experimentation clause where warranted.

# 7   Regulatory Learning as a Practice

Within anticipatory governance, regulatory learning goes hand in hand with experimentation and iterative policy making activities. in the spirit of anticipatory governance, it is vital to build in ex post evaluative review processes including feedback from stakeholders and where appropriate iterative tweaking to improve workability and efficacy and to take account of new and emerging insights. Iterative policymaking for innovation is succoured by this feedback loop provided by evaluative regulatory learning. Building regulatory learning into an anticipatory governance culture can deliver policy development and reform efficiencies. Anticipatory governance should avoid the path-dependent traditional approach of holding multiple, unfocused stakeholder consultations that regulators who are more timid around showing leadership on technological innovation may lean on, in the process missing the boat on delivering tangible iterative policy development and reflective regulatory learning as a practice. By contrast, anticipatory governance takes a leaf out of the innovator's playbook to engage in scaled-down real world testing of the regulatory scheme before its operationalisation or refinement. The suitability of far-seeing regulatory schemes can be road-tested to assess and refine their suitability before they are put into general operation. This evaluation with its attendant regulatory learning can help to future-proof the law by making it more relevant and robust.

Reflecting the spirit of iterative policymaking, changes of direction or policy through successive approximation arising from regulatory learning around innovation regulation ought not to be conceived of as mistakes. Rather they should be celebrated as contributing to the learning process. This coheres with Braithwaite's encouragement of the process of regulation to be carried out in a manner that is collaborative and conversational to 'build up a shared story book of successes and failures that constitutes shared sensibilities'.[151] Braithwaite speaks about errors as supplying fodder for learning opportunities about how to solve problems.[152] This is grist to the mill of iterative policymaking and regulation for innovation. In most cases the language of error or mistake is misconceived when discussing iterative policy development so long as the solution pursued was the best available one based on agreed collaborative and informed participatory foresight practices. Within anticipatory governance, arguably the only real regulatory policy failures that could arise from attempting to proactively address fast-moving technological innovation through dynamic policy approximation would be (i) failure to properly prepare using the available tools of anticipatory governance such as regulatory foresight; and (ii) failure to study the learning insights that the failure delivers and to use them to inform the next response or policy iteration. Of course not all outputs of anticipatory governance are policy instruments. It is entirely legitimate that the regulatory learning from a policy lab that is leveraging regulatory foresight at the leading-edge may be a conclusion that it remains too early to progress a policy agenda and that further research is required on the technological innovation and its application before there is a readiness to engage at a policy level. Nonetheless, future learning demands that an anticipatory governance culture should humbly reflect in order to retrospectively identify opportunity failings–instances of omission to develop policy options in a timely fashion to support anticipatory innovation.[153]

---

[151] Braithwaite (2008), p. 59.
[152] Braithwaite (2008), p. 59.

It needs to be emphasised that anticipatory governance and with it the era of meaningful regulatory learning as part of a continuous cycle of anticipatory governance and regulatory experimentation around innovation is still slowly dawning. It is early days in relation to harvesting the kind of learning that strategic regulatory foresight and measures such as regulatory sandboxes, pilot regulation, policy labs and experimentation clauses offer. However, as exemplified in the DLT Pilot regime discussed earlier,[154] some beacons are emerging of traditional and leisurely and top-down policymaking being bypassed in favour of an experimental, small-scale, evidence-based modelling approach to policy design that efficiently gathers evidence from the ground and evaluates it before deciding how to proceed. Within anticipatory governance, policymakers should also interrogate instrument choices such whether hard regulation is in fact an appropriate response to emergent or emerging innovation.[155] Adaptable rules and standards may be more suited to deftly handling such innovation rather than the blunt edge of legislation.

As regards regulatory learning becoming more embedded in regulatory ecology, the European Union's leadership and models may help Member States to make the shift to adapt their regulatory culture and processes for iteration and regulatory learning. The Joint Research Centre of the European Commission has led out on the use of policy labs at EU level and its work has provided regulatory learning that assists the European Commission in its policy agenda. An article co-authored by two members of staff at the European Commission's Joint Research Centre discussing the #Blockchain4EU project, examined its potential policy impacts as well as other dimensions such as economic, social, environmental and technical.[156] This showed that forward-looking physical prototypes developed for the #Blockchain4EU project to model scenarios and strategic policy recommendations around the future of blockchain of policy value, a fact acknowledged by the European Parliament in its Resolution on DLTs and blockchain.[157] Furthermore, the project's perspectives influenced the development of a funded initiative of the European Commission for development of blockchain applications for SMEs,[158] and separately the Commission's thinking on how blockchain could assist with transparent and resilient supply chains and with intellectual property rights.[159]

More formal EU regulatory learning from regulatory experimentation is set to come. The European Union plans to use insights gained from the 3 year DLT pilot when considering devising a more ambitious EU legislative framework for DLT in capital markets.[160] The use of the experiment for DLT developers and other stakeholders to provide broad, evidence-based future regulatory learning for legislative and supervisory purposes is made plain in the Recitals to the DLTPR which indicate that regulatory learning in relation to the opportunities and risks concerning crypto-assets as financial instruments and their underlying technologies will inform the question of making changes to the law to provide a fit for purpose regulatory framework. This shows the EU's commitment to expressly embracing regulatory learning as a reflective practice.

---

[153] Roberts (2018).
[154] DLT Pilot Regulation (EU) 2022/858.
[155] Emergent innovation is further back on the journey through the innovation lifecycle than emerging innovation.
[156] Pólova and Nascimento (2021).
[157] European Parliament (2018).
[158] European Commission. (2018b).
[159] European Commission (2017).
[160] European Commission (2023d), p. 18.

## 8  How Do We Evaluate Anticipatory Governance and its Impacts?

It must be acknowledged that there are a number of challenges to evaluating anticipatory governance and its costs and benefits. First, assessing the impact of foresight-related activities with hindsight is certainly premature when the system has not yet had sufficient time to become established. Currently the world of anticipatory governance is very much in its infancy, making assessment of how its performance problematic in the absence of (i) clear connections with outputs being capable of being transparently drawn, and (ii) independent assessment being carried out of anticipatory governance systems and agencies. Second, a core challenge in evaluating anticipatory governance is that it is resource-intensive but the intensity of the resources devoted does not correspond to linear, tangible policy outputs. In terms of a cost-benefit analysis, a regulatory foresight unit demands investing in people and other resources but as it is working with trendspotting and identifying and analysing potential future scenarios, its actual impacts in a concrete sense may be difficult to judge in the short to medium term.

Cost-benefit analysis can be a useful mechanism for assessing the welfare implications of policy or regulatory intervention. Any systematic cost-benefit analysis of anticipatory governance would need to consider the direct and indirect impacts of undertaking regulatory foresight and anticipatory governance activities. A key question is what benefits does an anticipatory governance culture deliver that would not otherwise occur? On one metric, the resources associated with anticipatory governance could be measured as a function of the outcomes that result. It can be difficult to state what the positive or beneficial effects will be on innovation, competition and productivity. We are essentially proceeding from the baseline scenario of what would happen or not happen in the absence of regulatory foresight activity and any follow-on consequences. This is difficult given the level of uncertainty involved in horizon scanning which means that identification of strong foresight signals indicating an associated need for intervention is not necessarily a given. Deciding when to regulate is far from an exact science but hesitating too long over devising a perfect regulatory scheme is perhaps the worst failing under anticipatory governance which is aimed not at perfection, but responsiveness. The real benefit of a regulatory foresight unit is the sense-making, transversal, cross-policy division advice it can give to policymakers, connecting the dots between a range of things on the horizon to empower both strategy and policy decisions.

Foresight activities involve a huge amount of sifting of information and assessing the strength of the signals. This is highly valuable but akin to panning for gold – the ex ante labour needed is considerable to yield a nugget that is worth following up on. This is worth doing even if the yield or resources is not high. Value for money-style political thinking may bring pressure to show tangible results such as policy outcomes. However, in any evaluation of what anticipatory governance has delivered in a given system, it would be a mistake to only give weight to evidence of actual policy developments around regulating new technologies and their applications. Within anticipatory governance, policymakers need to carefully consider when the right time to regulate is. This is all the more relevant with cutting-edge innovation. Within a cost-benefit analysis, applying the proportionality principle will often buttress the position that 'doing nothing', other than maintaining a watching brief for the time being, is the most credible policy option.[161] The relative strength or weakness of the signal obtained from foresight methodologies is relevant in making such a call in any case. Sometimes the right decision at the relevant point in time of the evolution of the innovation is that concentrating resources on learning and engaging with industry and stakeholders is all that is required so as to learn more about emergent uncertainties and plausible futures. This informed determination which anticipatory governance enables is an advance on a purely passive, wait and see approach.

---

[161] Regulatory Horizons Council (2022), p. 29.

When evaluating impact for anticipatory governance, it is tempting to want to point to a new regulatory initiative being prompted that enables commercialisation a new market for goods or services. That can sometimes occur. MiCA represents an early outlier in this respect. A market-enabling policy output is evident in the drive towards MiCA[162] beginning with the European Commission's Fintech Action Plan in 2018.[163] As Ferreira and Sandner correctly observe, this was about far more than filling regulatory gaps around the crypto industry – its objective was 'to successfully embed crypto assets into the fabric of the European economy to harvest the benefits of this innovation as much as it is to mitigate its potential risks.'[164] In the era of MICA, we can expect to start to see market impact as crypto-asset firms seek authorisation as credit institutions in the EU.[165] However, as already stressed, anticipatory governance impacts are not always tangible regulatory initiatives. Given the anticipatory nature of the activities often the benefits may simply be that policymakers are well briefed far in advance of any policy intervention being needed. These well-briefed decisionmakers then get to make the call on an informed basis as to whether resources should be devoted to the phenomenon and when. Activating an anticipatory governance culture means that opportunities can be catalysed through developing enabling regulatory frameworks and risks avoided through imposition of appropriate constraints. Missteps around determining how to optimally regulate technological advancements may be made in anticipatory governance culture, but iterative, regulatory experimentation in most cases is likely to trump the alternative of being outpaced through passive inaction. Beyond that, anticipated costs and benefits are complex to quantify.

Anticipatory governance clearly involves empowerment to recognise when it is time to be bold and imaginative. Having effective anticipatory governance mechanisms in place depends not just on superb foresight activities but also on a willingness for the system to grasp them with both hands and to constructively engage with the insights delivered to deliver regulation where this would be an apt response. This aspect distinguishes true and effective, joined-up anticipatory governance from a mere watching brief assignment. When the time is right, trialling bespoke regulatory frameworks for innovation with an effective regulatory learning feedback loop may lead to well-adapted frameworks facilitating the needs of nascent industries while also putting in place appropriate guardrails in place to protect the public. While earlier adoption of proportionate regulation for innovation than would otherwise be the case will likely impose compliance costs for participants in fledgling markets that increase the cost of going to market, on the other hand, timely bespoke regulation may give confidence and public trust to an emerging industry such as the clean technology sector. Complexities arise in that benefits of policy intervention do not fall uniformly: it is a truism that for every market player that can capitalise from a policy intervention, others may be negatively impacted as a consequence of new working method, business model, product or service coming on stream. Moreover, indirect impacts of regulation through differential compliance burdens can have significant consequences for competitive advantage.[166]

---

[162] Markets in Crypto-Assets Regulation (EU) 2023/1114.
[163] European Commission (2018a).
[164] Ferreira and Sandner (2021), p. 15.
[165] Enria (2023); Avgouleas and Seretakis (2023).
[166] Bartel and Thomas (1985); Dolar and Dale (2020).

One could measure anticipatory governance in terms of the opportunity costs associated with the alternative option not taken. Here while resources in terms of personnel and time would be saved by not embedding a systematic anticipatory governance culture, this would be counterbalanced by resultant knowledge gaps and policymakers not directing their attention to making timely, imaginative choices about what should happen next including deploying regulatory experimentation mechanisms. The associated regulatory delay in response could be considerable in terms of, not just time, but ambition in method and outcome. Regulatory foresight can bring innovative products to market more expeditiously and develop new markets such as the EU's bioeconomy.[167] But early intervention still requires taking a risk. Given that this is sophisticated crystal ball gazing, as the OECD observes, it is unsurprising that to date states and regions have largely focused their efforts on key sectors where there is an obvious strategic disadvantage to not being surprised – defence, cybersecurity, health and agriculture.[168] However, we are now at the point where much more is now demanded if private sector innovation is to be systematically enabled in a meaningful sense.

Moving the lens wider, there is a plethora of likely positive impacts for jurisdictions who are adaptive and embrace an anticipatory governance culture. Jurisdictions with an anticipatory governance culture are more likely to accrue reputational benefits as pro-innovation and to gain first mover advantages in attracting leading edge innovators. Anticipatory governance by regulators can thus confer strategic advantage on a jurisdiction in the eyes of innovators as compared with a jurisdiction that has not embraced it. Furthermore, anticipatory governance mechanism can help to consider how technologies can help to tackle grand challenges of our time and far-sighted policymaking can constitute a powerful driver of the flow of events including in relation to the green transition.

As yet systematic evaluations of anticipatory governance outcomes are rare. It is useful to consider the experience of the EU's Joint Research Centre which has led out as a pioneer on anticipatory governance, feeding into European policymaking. A report on an independent evaluation of its impact contains some useful insights into how its performance should be evaluated, including around foresight-driven policy work.[169] To help address the question of how do we evaluate the impact of foresight and anticipatory work, the independent evaluation recommended that the JRC gather data on its impact on regulation.[170] That seems reasonable. Of course policy responses short of formal regulation including pilots and experimentation may be apposite. It was envisaged in the review that the JRC should henceforth routinely adopt a proactive role in stimulating EU policymaking discussions, supported by its foresight work. A good case was made for a reinforced foresight and anticipation capacity to support the European Commission. It was suggested that this would 'make the JRC more effective in helping to shape the policy agenda and make the Commission's policy work more future proof.'[171]

While this approach may be accommodated at EU policymaking level, extrapolating to a national level, in the absence of a 'whole of government' approach which enhances joined up policymaking including an anticipatory governance culture, some resistance may be encountered by anticipatory governance actors in the form of path dependencies around how policy occurs and which actors have active policy making roles rather than briefing roles. This could impede the forward-oriented approach that is the essence of an anticipatory governance culture. The OECD articulates well the concern that anticipatory governance should not be subverted into being normal policymaking in that 'focusing on the overly-practical applications, making anticipatory work compete with business-as-usual activities, and not defending "out there" thinking can make anticipatory governance vulnerable to disruption or derailing its use.'[172]

---

[167] Carrez and Rupp (2023), p. 51.
[168] Roberts (2018).
[169] Heuer et al. (2022).
[170] Heuer et al. (2022), p. 25.
[171] Heuer et al. (2022), p 5.

It is appropriate to acknowledge that the culture of anticipatory governance for innovation envisaged here remains fresh rather than strongly embedded. At this early point in their evolution, we do not have an empirical study in relation to the correlation between regulatory experimentation projects, knowledge transfer and the impact on policy outcomes in innovation policy and regulation. This reflects the newness of anticipatory governance approaches in the policy landscape. Until more countries and regions have unequivocally grasped the nettle on anticipatory governance mechanisms, trialling regulation around innovation and transparently evaluating it, there is insufficient material for scholars to systematically evaluate. In time such a study would assist evaluation of anticipatory governance policy regulatory outcomes. This should become feasible and a concern for scholars assuming that mechanisms such as pilot regulation and policy labs become more widespread and entrenched and their influence on policymaking becomes common. The first step is for anticipatory governance to be more widely rolled out and embraced. Scepticism in the corridors of power can defeat the potential impact of an anticipatory governance culture. Certainly, while far-removed from crystal ball-gazing, future forecasting using regulatory foresight techniques require investment and will by definition not always be accurate. Nonetheless, the public interest demands that we engage with future-planning for innovation through probing plausible possible futures. The alternative (which is the status quo in many jurisdictions) is to soldier on with reactive governance that fails to take the long view and can be 'inattentive to its best options until they have been allowed to slide by.'[173] Making the required leap of faith is arduous but worthwhile.

The case for welcoming anticipatory governance for innovation is well-made. A well-functioning anticipatory governance culture should address head on the traditional pacing problem where regulatory stances either lag technological innovation or regulate too early. Fuelled by the advantage of vision front-loading in the form of regulatory foresight, policymakers not only keep pace with innovation but also anticipate what is on the horizon. They are apprised of what may be coming down the tracks, its level of technological maturity and the risks and opportunities that may present and use that to consider the optimal strategy to take. As systematic anticipatory governance beds down, adaptive, prescient policymaking should constitute the most tangible impact. In the process the trope dismantled by Mazzucato of the 'boring, lethargic State versus a dynamic private sector' will be dispelled.[174]

## 9    Conclusion

This article has sought to expand doctrinal debate and advance scholarship on innovation policy. Enabling the roll-out of responsible innovation in a timely fashion constitutes one of the most pressing challenges for jurisdictions who want to remain competitive in the innovation race. The new anticipatory governance equips states with the right tools to develop a timely, proportionate approach to innovation while proactively confronting the pacing problem and the Collingridge Dilemma head on. Such an agile ecology uniquely runs the gamut from trend and foresight analysis right up to collaborative policy design experimenting and trialling. Although the ingenious concept of anticipatory governance is very much new and emerging, it can be concluded that it has manifest potential to deliver an agile yet responsible regulatory ecology for innovation.

---

[172] Organisation for Economic Co-operation and Development (2020), p. 120.
[173] Fuerth (2009), p. 14.
[174] Mazzucato (2014), .p 4.

That said, anticipatory governance involves a challenging step change in regulatory approach. This article has focused on a variety of tools and foundations needed to build a strong anticipatory governance regulatory culture Three core elements have been framed here as forming the triumvirate of a robust and nimble anticipatory governance system – regulatory foresight, regulatory experimentation and regulatory learning. Furthermore, five guiding principles for agile anticipatory governance as a culture have been posited. More broadly, the new anticipatory governance culture must be underpinned by a paradigm shift in regulatory culture and as discussed here, an alignment of the right architecture for anticipatory governance with a public sector that is equipped to innovate and reflect. The impact of anticipation and experimentation will be augmented by a cohesive culture of networked anticipatory governance which is deeply reflective. It is also worth remembering that an anticipatory governance regulatory culture is a continuing, reflective journey rather than a destination.

This article has also sought to probe the complex question of how anticipatory governance should be evaluated and has identified a number of challenges. It would be a mistake to evaluate anticipatory governance solely by seeking solely to identify results that constitute high hanging fruit in the form of adopted regulation. Anticipatory governance is about so much more than that. Moreover, where they occur, policy interventions arrive at the furthest point not the first point in an anticipatory governance arc and at this point in the early emergence of anticipatory governance techniques, it is too soon to witness an entrenched impact of the tools of anticipatory governance on tangible policy outcomes. Within anticipatory governance, as forms of regulatory experimentation bed down, continuous improvement through evaluation and reflection upon regulatory learnings is crucial. This is the very essence of experimentation, regulatory learning and iterative policymaking. Indeed, a signifier of success will be that as a suitable regulatory policy approximation for applications of emerging technologies is reached, forms of regulatory experimentation such as regulatory sandboxes, policy labs or experimentation clauses become obsolete in that particular context.[175] Within the European Union, the DLT Pilot and the planned operation of regulatory sandboxes under the AI Act each represent major EU landmarks in the inculcation of regulatory experimentation techniques for technological innovation from the page to reality. Accordingly, their efficacy in practice will need to be scrutinised in due course.

Within the spirit of anticipatory governance, we must judge developments not only with the 360 degree perfection of hindsight, but also with foresight. The expectation is that governments that are forward-thinking around innovation and keen to attract innovators will be open to rework their policymaking structures to incorporate aspects of anticipatory governance culture so as to benefit from the transdisciplinary wisdom that comes from strategic regulatory foresight, policy labs and other forms of future scenario policy design and iterative policymaking. The Joint Research Centre constitutes a beacon for anticipatory governance activities from which other Member States can benefit and learn. Furthermore, the European Union's prescient role as flagbearer should hasten buy-in to anticipatory governance. Looking ahead it is likely that we will increasingly see formal drivers in the form of EU frameworks that enable both innovators and regulators to experiment, test and learn. To conclude, anticipatory governance may be new but the benefits of anticipating what comes next both to stay ahead of the innovation curve and to shape it are undeniable.

---

[175] This is sometimes legislatively recognised through the use of a sunset clause.

## References

Ahern D (2018) Regulatory arbitrage in a FinTech world: devising an optimal EU regulatory response to crowdlending. JBL 3: 193-214

Ahern D (2021) Regulatory lag, regulatory friction and regulatory transition as FinTech Disenablers: Calibrating an EU Response to the Regulatory Sandbox Phenomenon. EBOR 22:395–432. https://doi.org/10.1007/s40804-021-00217

Ahern D (2019) Regulators nurturing FinTech: global evolution of the regulatory sandbox as opportunity-based regulation. Indian Journal of Law and Technology 15: 345-378

Ahern D (2023) The role of sectoral regulators and other state actors in formulating novel and alternative pro-competition mechanisms in fintech. In: Stylianou K, Iacovides M, Lundqvist B (eds) Fintech competition: law, policy and market organisation. Bloomsbury-Hart Publishing, Oxford, pp 307–330. https://doi.org/10.5040/9781509963379.ch-012

Allen HJ (2019) Regulatory sandboxes. Geo Wash Law Rev 87: 579-645

Alonso Raposo M, Mourtzouchou A, Garys A, Brinkhoff-Button N, Kert K, Ciuffo B (2021) JRC future mobility solutions living lab: Conceptual framework, state of play and way forward. https://publications.jrc.ec.europa.eu/repository/handle/JRC127272. Accessed 12 March 2024

Armstrong H, Gorst C, Rae J (2019) Renewing regulation: 'anticipatory regulation' in an age of disruption. Nesta, UK

Arner DW, Zetzche DA, Buckley RP, Kirkwood JM (2023) The financialisation of crypto. European Banking Institute Working Paper No. 148 https://ebi-europa.eu/publication-working-paper-series-no-148/

Arntzen S, Wilcox Z, Lee N, Hadfeld C, Rae J (2019) Testing innovation in the real world: real-world testbeds. Nesta, UK

Avgouleas E and Seretakis A (2023) How should crypto lending be regulated under EU law? EBIOR 23:421-438. https://doi.org/10.1007/s40804-023-00293-3

Bailey J, Hinrichs-Krapels S, Pollitt A, Duffy B (2019) Policy and Innovation Lab Landscape Review. Technical Internal Document. The Policy Institute, King's College London

Baldwin R (2010) Better regulation: the search and the struggle. In: Baldwin R, Cave M, Lodge M (eds) The Oxford handbook of regulation. Oxford University Press, Oxford, New York, pp 259-278

Bartel AP, Thomas LG (1985) Direct and indirect effects of regulation: A new look at OSHA's impact. Journal of Law and Economics 28:1-25

Baumgart M, Lavrijssen S (2023) Exploring regulatory strategies for accelerating the development of sustainable hydrogen markets in the European Union. Journal of Energy and Natural Resources Law. https://doi.org/10.1080/02646811.2023.2257528

Bolton M, Mintrom M (2023) RegTech and creating public value: Opportunities and challenges. Policy Design & Practice 6:266-282. https://doi.org/10.1080/25741292.2023.2213059

Braithwaite J (2005) Markets in Vice, Markets in Virtue . Oxford and Federation Press, New York and Sydney

Braithwaite J (2008) Regulatory Capitalism. Edward Elgar, Cheltenham

Bunea A, Chrisp J (2022) Reconciling participatory and evidence-based policymaking in the EU Better regulation policy: mission (im)possible? Journal of European Integration 45:729–750

Buocz T, Pfotenhauer S and Eisenberger I (2023) Regulatory sandboxes in the AI Act: reconciling innovation and safety? 15:357-389. https://doi.org/10.1080/17579961.2023.2245678

Cagnin C, Havas A, Saritas O (2013) Future-oriented technology analysis: Its potential to address disruptive transformations. Technological Forecasting and Social Change 80: 379-385. https://doi.org/10.1016/j.techfore.2012.10.001

Carrez D, Rupp M (2023) A thriving bioeconomy needs better tools for measuring and monitoring. EuroChoices 22:51-55. https://doi.org/10.1111/1746-692X.12411

Centre for Regulatory Innovation Canada (2023) What is regulatory experimentation? https://www.canada.ca/en/government/system/laws/developing-improving-federal-regulations/modernizing-regulations/regulatory-experimentation.html. Accessed 12 March 2024

Clements R (2023) Entry barriers in fintech. In: Stylianou K, Iacovides M, Lundqvist B (eds) Fintech competition: law, policy and market organisation. Bloomsbury-Hart Publishing, Oxford, pp 25–46. https://doi.org/10.5040/9781509963379.ch-002

Collingridge D (1980), The social control of technology. Pinter, London

Department for Science, Innovation and Technology (2024) Consultation outcome - A pro-innovation approach to AI regulation: government response Command Paper CP 2019 https://www.gov.uk/government/consultations/ai-regulation-a-pro-innovation-approach-policy-proposals/outcome/a-pro-innovation-approach-to-ai-regulation-government-response

Dolar B, Dale B (2020) The Dodd-Frank Act's non-uniform regulatory impact on the banking industry. Journal of Banking Regulation 21:188-195. https://doi.org/10.1057/s41261-019-00106-z

Drahos P, Krygier M (2017) Regulations, institutions and networks. In: Drahos P (ed), Regulatory theory: foundations and applications, ANU Press, Canberra, pp.1-22

ElZarrad MK, Lee AY, Purcell R, Steele SJ (2022) Advancing an agile regulatory ecosystem to respond to the rapid development of innovative technologies. Clinical and Translational Science 15:1332–1339 https://doi.org/10.1111/cts.13267

Enria A (2023) 'Regulating crypto finance: taking stock and looking ahead' Speech of the Chair of the Supervisory Board of the ECB, Venice, 14 November 2023 https://www.bankingsupervision.europa.eu/press/speeches/date/2023/html/ssm.sp231114~fd1b2cc234.en.html

European Commission (2018a), Communication from the Commission to the European Parliament, the Council, the European Central Bank, the European Economic and Social Committee and the Committee of the Regions: FinTech Action Plan: For a more competitive and innovative European financial sector COM/2018.0109

European Commission (2018b). Blockchain and distributed ledger technologies for SMEs (INNOSUP-03-2018)

European Commission (2020) A digital finance strategy for the EU. COM/2020/591

European Commission (2021) Proposal for a Regulation of the European Parliament and of the Council laying down harmonised rules on Artificial Intelligence (Artificial Intelligence Act) and amending certain Union legislative acts. COM(2021) 206 final. https://eur-lex.europa.eu/legal-content/EN/TXT/?uri=celex%3A52021PC0206. Accessed 12 March 2024

European Commission (2022a) Proposal for a Regulation of the European Parliament and of the Council laying down measures for a high level of public sector interoperability across the Union (Interoperable Europe Act) COM (2022) 720 final Brussels 18.11.2022

European Commission (2022b) Proposal for a Directive of the European Parliament and of the Council amending Directive 2010/75/EU of the Parliament and of the Council of 24 November 2010 on industrial emissions (integrated pollution prevention and control) and Council Directive 1999/31/EC of 26 April 1999 on the landfill of waste COM/2022/156

European Commission (2022c) First regulatory sandbox on Artificial Intelligence presented 27 June 2022 https://digital-strategy.ec.europa.eu/en/news/first-regulatory-sandbox-artificial-intelligence-presented. Accessed 12 March 2024

European Commission (2023a) AI Pact. https://digital-strategy.ec.europa.eu/en/policies/ai-pact. Accessed 12 March 2024

European Commission (2023b) Better regulation: why and how. https://commission.europa.eu/law/law-making-process/planning-and-proposing-law/better-regulation_en. Accessed 12 March 2024

European Commission (2023c) Communication from the Commission to the European Parliament, the Council, the European Economic and Social Committee and the Committee of the Regions: An EU initiative on web 4.0 and virtual worlds: a head start in the next technological transition. COM(2023) 442 final. https://digital-strategy.ec.europa.eu/en/library/eu-initiative-virtual-worlds-head-start-next-technological-transition. Accessed 12 March 2024

European Commission (2023d) Regulatory learning in the EU: Guidance on regulatory sandboxes, testbeds, and living labs in the EU, with a focus section on energy SWD(2023) 277/2 final https://research-and-innovation.ec.europa.eu/system/files/2023-08/swd_2023_277_f1.pdf. Accessed 12 March 2024

European Commission (2023e) SME relief package 12 September 2023 https://single-market-economy.ec.europa.eu/system/files/2023-09/COM_2023_535_1_EN_ACT_part1_v12.pdf. Accessed 12 March 2024

European Commission (2023f) Strategic foresight report 2023 https://commission.europa.eu/system/files/2023-07/SFR-23-beautified-version_en_0.pdf. Accessed 12 March 2024

European Commission (2024), European blockchain sandbox best practices report (2023). https://digital-strategy.ec.europa.eu/en/library/european-blockchain-sandbox-best-practices-report. Accessed 12 March 2024.

European Council (2020) Conclusions on regulatory sandboxes and experimentation clauses as tools for an innovation-friendly, future-proof and resilient regulatory framework that masters disruptive challenges in the digital age 2020/C 447.01 OJ C 447 23.12.2020, p 1-3. https://eur-lex.europa.eu/legal-content/EN/TXT/?uri=CELEX%3A52020XG1223%2801%29. Accessed 12 March 2024

European Innovation Council (2023) EIC tech report 2023 https://eic.ec.europa.eu/system/files/2023-10/EIC-TechReport-2023-DigitalVersion_0.pdf. Accessed 12 March 2024.

European Parliament (2018). Resolution on DLTs and blockchains: Building trust with disintermediation (2017/2772(RSP)). http://www. europarl.europa.eu/sides/getDoc.do?type=MOTION&reference=B8-2018-0397&format=XML&language=EN. Accessed 12 March 2024

European Securities and Markets Authority, Questions and answers on the implementation of Regulation (EU) 2022/858 of the European Parliament and of the Council of 30 May 2022 on a pilot regime for market infrastructures based on distributed ledger technology 2 June 2023 ESMA70-460-189 https://www.esma.europa.eu/sites/default/files/library/esma70-460-189_qas_dlt_pilot_regulation.pdf. Accessed 12 March 2024

Federal Ministry for Economic Affairs and Energy (Germany) (2020) New flexibility for innovation: guide for formulating experimentation clauses. https://www.bmwk.de/Redaktion/EN/Publikationen/Digitale-Welt/guide-new-flexibility-for-innovation-en-web-bf.pdf?__blob=publicationFile&v=1. Accessed 12 March 2024.

Fenwick M, Vermeulen EPM (2020) Fintech, overcoming friction and new models of regulation. In: Fenwick M, Van Uytsel S, Ying B (eds) Regulating FinTech in Asia. Perspectives in Law, Business and Innovation. 2020 Springer, Singapore, pp 205-225

Ferreira A, Sandner P (2021) EU search for regulatory answer to crypto assets and their place in the financial markets' infrastructure. Computer Law & Security Review 43: 105632., 1-15. https://doi.org/10.1016/j.clsr.2021.105632

Financial Conduct Authority (United Kingdom) (2023a) Regulating cryptoassets: Phase 1: Stablecoins. Discussion Paper DP23/4. https://www.fca.org.uk/publication/discussion/dp23-4.pdf. Accessed 15 February 2024

Financial Conduct Authority (United Kingdom) (2023b) Emerging technology research hub. https://www.fca.org.uk/firms/emerging-technology-research-hub. Accessed 15 February 2024

France Expérimentation (2023a) Faire évoluer le droit avec ceux qui font l'innovation. https://www.modernisation.gouv.fr/files/2023-05/France%20Exp%C3%A9rimentation-brochure_0.pdf . Accessed 12 March 2024

France Expérimentation (2023b) Direction interministérielle publique de la transformation publique https://www.modernisation.gouv.fr/transformer-laction-publique/france-experimentation. Accessed 12 March 2024

Fuerth LS (2009) Foresight and anticipatory governance. Foresight 11:14–32. https://doi.org/10.1108/14636680910982412

Fuerth LS, Faber EMH (2013) Anticipatory governance: Winning the future. The Futurist 47:42–49

Gangale F, Mengolini, A, Covrig, L, Chondrogiannis S, Shorthall R (2023) Making energy regulation fit for purpose – State of play of regulatory experimentation in the EU – Insights from running regulatory sandboxes. Joint Research Centre of the European Commission, Publications Office of the European Union. https://data.europa.eu/doi/10.2760/32253. Accessed 12 March 2024

Gerrans P, Baur DG, Lavagna-Slater S (2022), Fintech and responsibility: Buy-now-pay-later arrangements. Australian Journal of Management 47:474-502. https://doi.org/10.1177/03128962211032448

Gluckman PD, Bardsley A, Kaiser M (2021) Brokerage at the science-policy interface: from conceptual framework to practical guidance. 8 humanities and social sciences communications 84:1-10. https://doi.org/10.1057/s41599-021-00756-3

Gofen A, Golan E (2020) Laboratories of design: a catalog of policy innovation labs in Europe. Hebrew University of Jerusalem. http://dx.doi.org/10.2139/ssrn.3822821

Government Office for Science (UK) (2017) The futures toolkit: tools for futures thinking and foresight across UK Government https://assets.publishing.service.gov.uk/media/5a821fdee5274a2e8ab579ef/futures-toolkit-edition-1.pdf. Accessed 12 March 2024

Grabosky P (2017) Meta-regulation. In: Drahos P (ed), Regulatory theory: foundations and applications. ANU Press, Canberra, pp 149-153

Guston DH (2014) Understanding 'anticipatory governance'. Social Studies of Science, 44:218-242. https://doi.org/10.1177/0306312713508669

Haines F (2017) Regulation and risk. In: Drahos P (ed), Regulatory theory: foundations and applications. ANU Press, Canberra, pp 181-196

Hebinck A, Vervoort JM, Hebinck P, Rutting L, Galli F (2018) Imagining transformative futures: Participatory foresight for food systems change. Ecology and Society, 23:2 https://doi.org/10.5751/ES-10054-230216

Heuer RD et al. (2022) Ex post evaluation of the activities of the Joint Research Centre under Horizon 2020 and Euratom 2014-2020. JRC130367. Publications Office of the European Union, Luxembourg

Hinrichs-Krapels S, Bailey J, Boulding H, Duffy B, Hesketh R, Kinloch E, Pollitt A, Rawlings S, van Rij A, Wilkinson B, Pow R, Grant G (2020) Using policy labs as a process to bring evidence closer to public policymaking: A guide to one approach. 6 Palgrave Communications 6: 101 https://doi.org/10.1057/s41599-020-0453-0

Joint Research Council (2022), JRC strategy 2030 https://joint-research-centre.ec.europa.eu/system/files/2022-12/JRC%20Strategy%202030.pdf. Accessed 12 March 2024

Kert K, Vebrova M, Schade S (2022) Regulatory learning in experimentation spaces. JRC130458 Joint Research Centre, European Commission. https://publications.jrc.ec.europa.eu/repository/handle/JRC130458.

Kimbell L, Bailey J (2017) Prototyping and the new spirit of policy-making. CoDesign, 13: 214–226. https://doi.org/10.1080/15710882.2017.1355003

Kingdon JW (2003) Agendas, alternatives and public policies. 2nd edition, Longman, New York

Kwek J, Parkash, SG (2020) Strategic foresight: How policymakers can make sense of a turbulent world. Apolitical. https://apolitical.co/solution-articles/en/strategic-foresight-making-sense-of-a-turbulent-world. Accessed 12 March 2024

Lauber K, Brooks E (2023) Why meta-regulation matters for public health: The case of the EU better regulation agenda. Globalization and Health, 19:1-70. https://doi.org/10.1186/s12992-023-00971-4

Lee C, Ling M (2020) The role of policy labs in policy experiment and knowledge transfer: A comparison across the UK, Denmark, and Singapore. Journal of Comparative Policy Analysis 22:281-29. https://doi.org/10.1080/13876988.2019.1668657

Macnaghten P (2021) Towards an anticipatory public engagement methodology: deliberative experiments in the assembly of possible worlds using focus groups. Qualitative Research 21:3–19. https://doi.org/10.1177/1468794120919096

McAleavey P (2023) Revitalising the JRC strategy 2030. Publications Office of the European Union, Luxembourg

McGann M, Blomkamp E, Lewis JM (2018) The rise of public sector innovation labs: Experiments in design thinking for policy. Policy Sciences 51:249-267. https://doi.org/10.1007/s11077-018-9315-7

McNulty D, Miglionico A, Milne A (2022) Technology and the 'new governance' techniques of financial regulation. European Banking Institute Working Paper Series No. 118. https://ebi-europa.eu/publication-working-paper-series-no-118/.

Maume P (2023) The regulation on markets in crypto-asses (MiCAR): Landmark codification, or first step of many, or both? ECFR 2:243-275. https://doi-org.eui.idm.oclc.org/10.1515/ecfr-2023-0014

Mazzucato M (2014) The entrepreneurial State. 2nd edition. Public Affairs. New York

Micheler E, Whaley A (2020) Regulatory technology: Replacing law with computer code. Eur Bus Org Law Rev 21: 349–377. https://doi-org.eui.idm.oclc.org/10.1007/s40804-019-00151-1

Miglionico A (2022) Regulating innovation through digital platforms: the sandbox tool. ECFR 19: 828-853. https://doi-org.eui.idm.oclc.org/10.1515/ecfr-2022-0029

Mintrom M, Luetjens J (2016) Design thinking in policymaking processes: Opportunities and challenges. Australian Journal of Public Administration 75:391–402

Morgan B (2003) The economization of politics: Meta-regulation as a form of nonjudicial legality. Soc Leg Stud 12:489-523

Muiderman K, Gupta A, Vervoot J, Biermann F (2020) Four approaches to anticipatory climate governance: Different conceptions of the future and implications for the present. WIRE's Climate Change e673 1-20 https://wires.onlinelibrary.wiley.com/doi/pdf/10.1002/wcc.673

Ojha H, Maheshwari B, Bhattarai B (2021) Co-creating knowledge, policy and practice: a call to advance water policy lab process. World Water Policy 7:132–142. https://doi.org/10.1002/wwp2.12050

Organisation for Economic Co-operation and Development (2021) Foresight and anticipatory governance in practice: Lessons in effective foresight institutionalisation. Strategic Foresight Office, Office of the Secretary General. https://www.oecd.org/strategic-foresight/ourwork/Foresight_and_Anticipatory_Governance.pdf. Accessed 12 March 2024

Organisation for Economic Co-operation and Development (2021). 'Overcoming exclusion in digital economies' in Development co-operation report 2021: Shaping a just digital transformation.

Organisation for Economic Co-operation and Development (2019) Strategic foresight for better policies: Building effective governance in the face of uncertain future. https://www.oecd.org/strategic-foresight/ourwork/Strategic%20Foresight%20for%20Better%20Policies.pdf. Accessed 12 March 2024

Organisation for Economic Co-operation and Development (2020) Regulatory policy in the Slovak Republic: Towards future-proof regulation. https://doi.org/10.1787/ce95a880-en.

Organisation for Economic Co-operation and Development (2023) What is strategic foresight? Webpage. https://www.oecd.org/strategic-foresight/. Accessed 12 March 2024

Parker C, Braithwaite J (2003) Regulation. In: Cane P, Tushnet M (eds), The Oxford handbook of legal studies. Oxford University Press, Oxford

Parsons W (2002) From muddling through to muddling up—Evidence based policy making and the modernisation of British government. Public Policy and Administration 17:43–60

Pollman E (2019) Tech, regulatory arbitrage and limits. Eur Bus Org Law Rev 20:567-590. https://doi.org/10.1007/s40804-019-00155-x

Pólova A, Nascimento S (2021) Foresight and design fictions meet at a policy lab: an experimentation approach in public sector innovation. Futures 128:102709. https://doi.org/10.1016/j.futures.2021.102709

Prosser JAW (2006) Regulation and Social Solidarity. Journal of Law and Society 33:364-387

Ranchordás S (2021) Experimental regulations and regulatory sandboxes – law without order? Law and Method. https://doi.org/10.5553/REM/.000064

Regulatory Horizons Council (UK) (2022) 'Closing the gap': Getting from principles to practices for innovation friendly regulation. https://assets.publishing.service.gov.uk/media/62ab5a668fa8f5356c35bb61/closing-the-gap-regulation-full-report.pdf. Accessed 12 March 2024.

Roberts A (2018) Innovation facets part 6: Anticipatory innovation. Blogpost 14 November 2018. Organisation for Economic Co-operation and Development. https://oecd-opsi.org/innovation-facets-part-6-anticipatory-innovation/. Accessed 12 March 2024

Rübig P (2012) The changing face of risk governance: Moving from precaution to smarter regulation' European Journal of Risk Regulation 3: 145-146. https://doi:10.1017/S1867299X00001975

Schumpeter JA (2008) Capitalism, socialism and democracy. 3rd edn, Harper Perennial, New York

Schuurman D, Tõnurist P (2017) Innovation in the public sector: Exploring the characteristics and potential of living labs and innovation labs. Technology Innovation Management Review 9:7–14

Secretary of State for Science, Innovation and Technology (UK) (2023) A pro-innovation approach to AI regulation Cmnd 815 (2023) https://www.gov.uk/government/publications/ai-regulation-a-pro-innovation-approach/white-paper. Accessed 12 March 2024

Seifert A (2020) The digital exclusion of older adults during the COVID-19 pandemic. Journal of Gerontological Social Work 63:674-676.

Selwyn N (2002) 'E-stablishing' and inclusive society? Technology, social exclusion and UK government policy making. Journal of Social Policy 31:1-20

Soeteman-Hernández LG, Sutcliffe HR, Sluijters T, van Guens J, Norrlander CW, Sips A J A M (2021) Modernizing innovation governance to meet policy ambitions through trusted environments. NanoImpact 21:100301. https://doi.org/10.1016/j.impact.2021.100301

Sparrow M (2000) The regulatory craft: Controlling risks, managing problems and managing compliance. The Brookings Institution, Washington DC

Strengers Y, Dahlgren K, Pink S, Sadowski J, Nicholls L (2022) Digital technology and energy imaginaries of future home life: Comic-strip scenarios as a method to disrupt energy industry futures. Energy Research & Social Science 84:102366. https://doi.org/10.1016/j.erss.2021.102366

Teubner G (1998) Juridification: Concepts, Aspects, Limits, Solutions. Oxford University Press, Oxford

Tõnurist P, Hanson A (2020) Anticipatory innovation governance: Shaping the future through proactive policy making .OECD working papers on public governance No.44. https://doi.org/10.1787/cce14d80-en

Topp L, Mair D, Smillie L, Cairney P (2018) Knowledge management for policy impact: the case of the European Commission's Joint Research Centre. Palgrave Commun 4:87. https://doi.org/10.1057/s41599-018-0143-3

Zilgavlis P (2014) The need for an innovation principle in regulatory impact assessment: the case of finance and innovation in Europe. Policy & Internet 6:377-392